\pdfoutput=1
\documentclass[aps, reprint, superscriptaddress, nofootinbib,preprintnumbers,twocolumn,showpacs,floatfix]{revtex4-1}
\usepackage{amsmath, latexsym, amssymb, graphicx, color, slashed, hyperref}
\definecolor{nicered}{rgb}{.7,.1,.1}
\definecolor{nicegreen}{rgb}{.1,.5,.1}
\definecolor{darkblue}{rgb}{0,0,.5}
\hypersetup{colorlinks, citecolor=nicegreen,linkcolor=nicered, urlcolor=darkblue}
\usepackage[T1]{fontenc}

\begin{document}

\title{Multifield Polygonal Bounces}

\author{Victor Guada}
\email{victor.guada@ijs.si}
\affiliation{Jo\v zef Stefan Institute, Jamova 39, Ljubljana, Slovenia}

\author{Alessio Maiezza}
\email{alessio.maiezza@irb.hr}
\affiliation{Ru\dj er Bo\v{s}kovi\'c Institute, Bijeni\v{c}ka cesta 54, Zagreb, Croatia}

\author{Miha Nemev\v{s}ek}
\email{miha.nemevsek@ijs.si}
\affiliation{Jo\v zef Stefan Institute, Jamova 39, Ljubljana, Slovenia}

\date{\today}

\begin{abstract}
We propose a new approach for computing tunneling rates in quantum or thermal field theory with multiple
scalar fields. It is based on exact analytical solutions of piecewise linear potentials with many
segments that describes any given potential to arbitrary precision. The method is first developed for the
single field case in 3 and 4 space-time dimensions and demonstrated on examples of classical potentials as
well as the calculation of quantum fluctuations. A systematic expansion of the potential beyond the linear
order is considered, taking into account higher order corrections, which paves the way for multiple scalar
fields. We thereby provide a fast semi-analytical tool for evaluating the bounce action for theories with an
extended scalar sector.
\end{abstract}


\maketitle


\section{Introduction}
Stability of the vacuum and phase transitions in the early universe are subjects of deep interest
to particle physics and cosmology. The non-perturbative problem of the tunneling among two vacua
was developed in seminal works~\cite{Kobzarev:1974cp, Coleman:1977py, Callan:1977pt, Coleman:1985}
for single scalar field theories. The problem of evaluating the lifetime of such metastable states was solved by
computing the action of a semiclassical instanton solution, called the bounce, interpolating between
the two minima. The form of the bounce was proven to have \(O(D)\) invariance under general
conditions~\cite{Coleman:1977th} for \(D>2\) in flat spacetime.

While finding the bounce in four dimensions is needed to assess the stability of the vacuum, an
analogous calculation becomes important at finite temperature. The bounce action
in \( D = 3 \) dimensions sets the probability of bubble nucleation~\cite{Linde:1980tt} and controls
the quality of the contingent phase transitions. Moreover, the shape of the field solution, e.g. the
size and thickness of the bubble is directly related to the power spectrum of gravitational
waves~\cite{Witten:1984rs, Hogan:1986qda} (see \cite{Cutting:2018tjt} for a recent analysis).

Computing the bounce action involves solving a nonlinear second order differential equation with
a friction term dependent on \(D\). Finding an analytical solution in a closed form is in general impossible for
an arbitrary potential. However, an approximation can be found in the thin-wall regime~\cite{Coleman:1977py}
and examples of exactly soluble potentials include the binomial, logarithmic~\cite{Aravind:2014pva} and 
quartic one~\cite{Lee:1985uv}. In most occasions the calculation of the bounce is thus performed numerically. For renormalizable 
single field potentials, one can use rescaling to define a single parametric problem and solve it by the usual
shooting method~\cite{Adams:1993zs, Sarid:1998sn}. Moreover, it is possible to derive an absolute lower bound
on the bounce action~\cite{Dasgupta:1996qu, Aravind:2014aza, Sato:2017iga, Brown:2017cca} and to provide
estimates based on a tunneling potential~\cite{Espinosa:2018hue} as well as machine learning techniques~\cite{Jinno:2018dek}.

A remarkably simple example of a soluble bounce is the linear potential. This is the basis
for our discussion that builds on the work of Duncan and Jensen (DJ)~\cite{Duncan:1992ai} in which
two linear segments are combined into a triangular potential barrier. The shooting is
transformed into an algebraic problem that is solved analytically in \(D = 4\).
This approximation was studied in~\cite{Dutta:2012qt} for single field
and~\cite{Masoumi:2017trx} for multi-field potentials. Another combination of two segments, one with a linear and
other with a quartic potential was considered in~\cite{Dutta:2011rc}. The analytical continuation of the
triangular solution from Euclidean to Minkowski space was developed in~\cite{Pastras:2011zr}.

Finding the Euclidean action becomes harder when an arbitrary number of fields is considered. As shown
recently~\cite{Blum:2016ipp}, the bounce still keeps the \(O(D)\) invariance. Nevertheless, finding
the path in field space and computing the bounce with multi-field potentials is significantly more
challenging. The main difficulty with the usual shooting approach is finding the fine-tuned initial field value
in the multidimensional field space, especially close to the thin wall limit, and integrating the system of 
coupled differential field equations.

There exist numerous approaches to the problem of multi-field tunneling. These include an improved action
method that converts the saddle point into a minimum~\cite{Kusenko:1995jv}, numerical functional
minimization~\cite{John:1998ip}, path deformation and shooting~\cite{Cline:1999wi, Wainwright:2011kj},
frictionless dimensional continuation~\cite{Konstandin:2006nd, Park:2010rh}, semi-analytical
techniques~\cite{Akula:2016gpl}, multiple shooting~\cite{Masoumi:2016wot}, tunneling
potential~\cite{Espinosa:2018szu} and numerically solving coupled PDEs with variable
coefficients~\cite{Athron:2019nbd}.

In this work we propose a new approach to obtain the bounce solution that is based on a generalization
of the DJ calculation, namely gluing an arbitrary number of linear segments into a polygonal potential
and solving the resulting system for any \(D\) and any number of fields. By increasing the number of
segments one can approximate any potential that admits a bounce solution, with an arbitrary precision
and thereby obtain the relevant action. The polygonal method enables one to work out bounces within
non-analytic potentials, even when the usual approaches may have issues with stability.

In \S\ref{secConstruct} we review the basics of vacuum tunneling, introduce the polygonal method
and construct the single field bounce solution.
We discuss how this approach is employed in \S\ref{secEvaluate}, where the relative convergence
is evaluated on selected problems and contact is made with the existing tools.
In \S\ref{secFluctuate} we show how these bounce solutions are used in the calculation of the decay
rate pre-factor from one loop quantum fluctuations.
In \S\ref{secExtend} we extend the method beyond the linear approximation and pave the way for
\S\ref{secMultiField}, where the multi-field case is developed.
We conclude with an outlook in \S\ref{secConclude} and leave details to appendices:
dimensions other than \(D=3,4\) are covered in~\ref{appD268}, the two segment calculation is expanded
in~\ref{appN3D} and further details on root finding can be found in~\ref{appRootFReR}.

%
%
\section{Single field polygonal bounces} \label{secConstruct}

\begin{figure}[!ht]
  \centering
  \includegraphics[width=.5 \columnwidth]{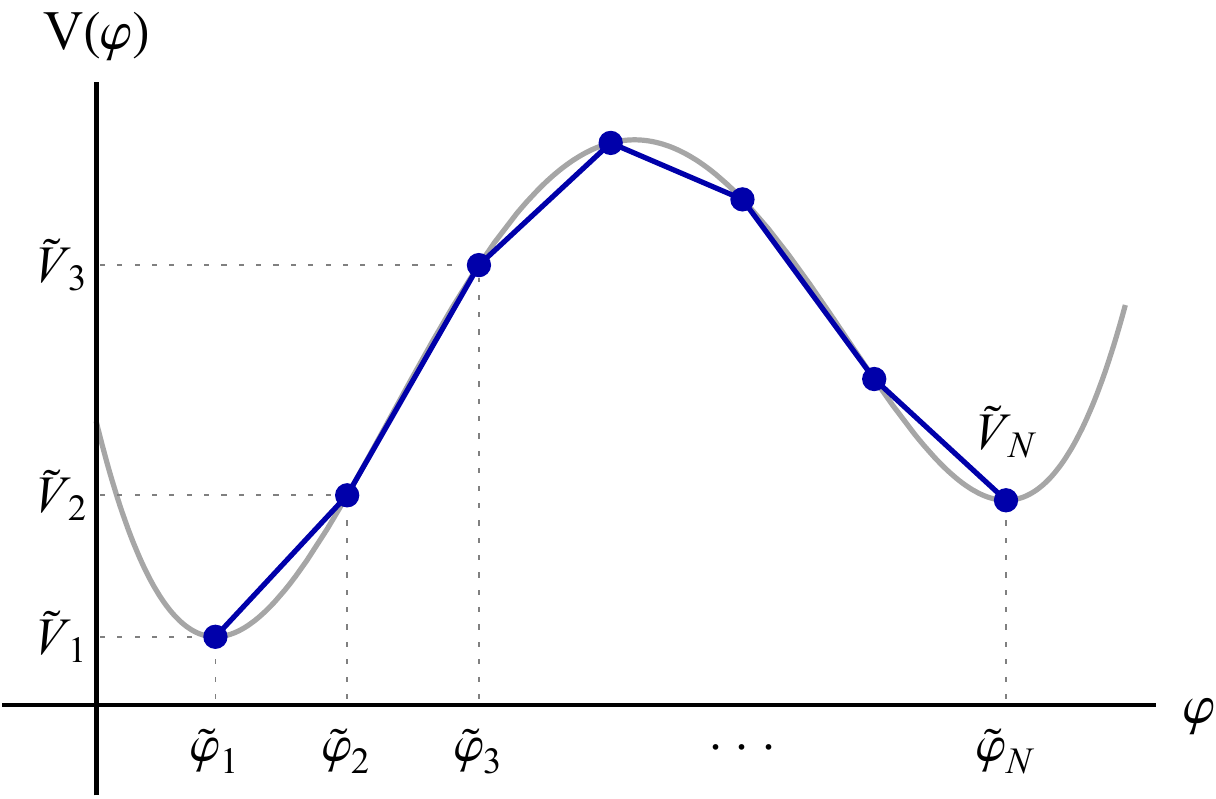}%
  \includegraphics[width=.5 \columnwidth]{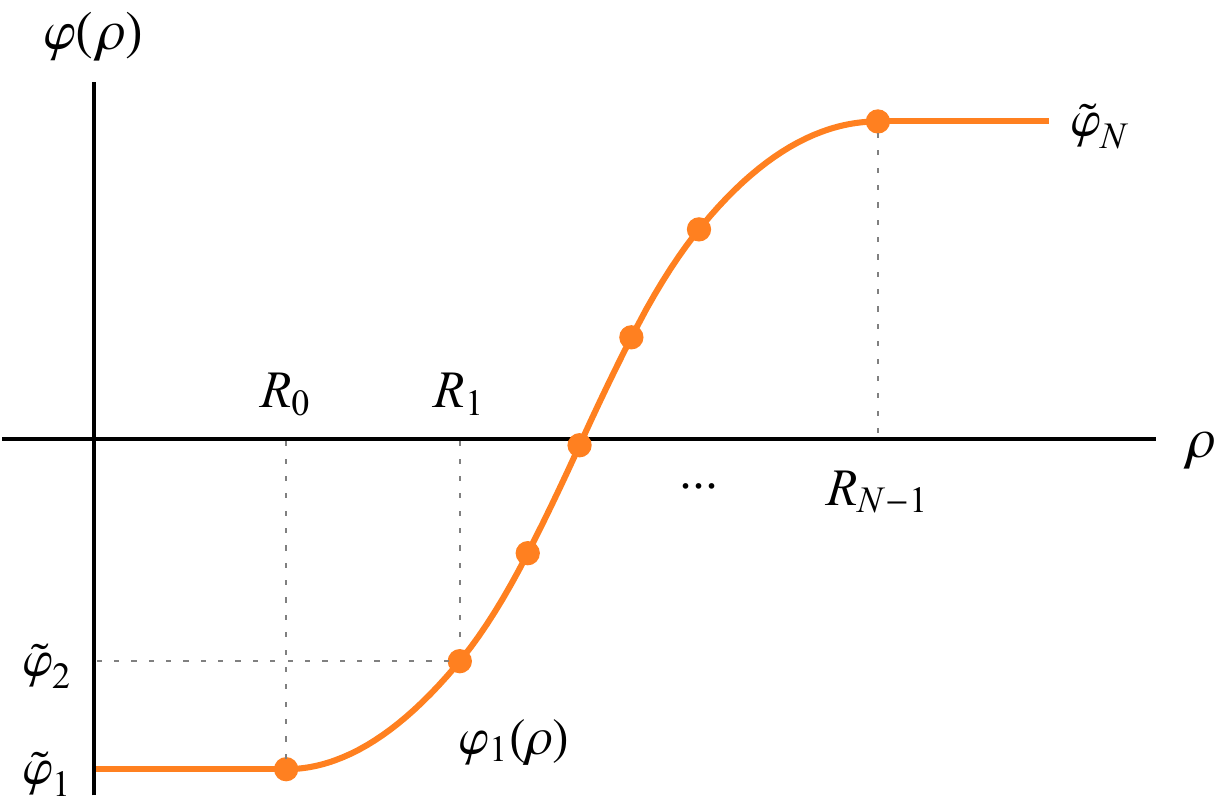}%
  \caption{
    Left: Linearly off-set quartic potential in gray and the polygonal approximation with \(N=7\) in blue.
    Right: The bounce field configuration corresponding to the potential on the left, computed with the
    polygonal bounce approximation.}
  \label{fig:VPhi}
\end{figure}

%
%
\subsection{Bounce redux}
Let us recall the basic features of vacuum transitions in field theory. We consider a
single real scalar field \(\varphi\) in \(D\) dimensions, subject to an arbitrary
potential \(V(\varphi)\) with non-degenerate minima, shown on the left of FIG.~\ref{fig:VPhi}.

The probability of tunneling from one ground state to another is proportional to the Euclidean
action \(S_D\). We assume the \(D\) dimensional solution to be \(O(D)\) symmetric~\cite{Coleman:1977th}
for any number of fields~\cite{Blum:2016ipp}
\begin{equation} \label{eqSED}
  S_D = \frac{2 \pi^{\frac{D}{2}}}{\Gamma \left(\frac{D}{2}\right)} \int_0^\infty \rho^{D-1}
  \text{ d} \rho \left(\frac{1}{2} \sum_i \dot \varphi_i^2 + V(\varphi_i)  \right),
\end{equation}
where \(\rho^2 = t^2 + \sum x_i^2 \) is the Euclidean radius that sets the size of the bubble.

The bounce is an instanton solution of the Euler-Lagrange equation that interpolates between the minima of \(V\)
and therefore obeys the appropriate boundary conditions
\begin{align} \label{eqBounceEqD}
  &\ddot \varphi_i + \frac{D-1}{\rho} \, \dot \varphi_i = d_i V, \nonumber \\
  &\varphi_i(0) = \varphi_{i 0}, \quad \varphi_i(\infty) = \tilde \varphi_{i N}, \quad \dot \varphi_i(0, \infty) = 0,
\end{align}
where \(d_i V\) is the derivative of \(V\) with respect to \(\varphi_i\). The field starts at \( \varphi_{i 0} \) with zero
velocity and rolls down to a stop in the false vacuum \( \tilde \varphi_{i N} \) at \( \rho = \infty \).

The usual shooting procedure involves numerically integrating the bounce Eq.~\eqref{eqBounceEqD} and varying
$\varphi_{i 0}$ until the boundary conditions are met. In this procedure, care should be taken when numerically
evaluating \(\rho \to 0 \) or \(\infty\). Conversely, notions of zero and infinity are not relevant for polygonal bounces below.

%
%
\subsection{Polygonal bounces}
In this work we introduce the polygonal bounce (PB) by generalizing the approach of~\cite{Duncan:1992ai}.
Instead of the generic potential with two minima, let \(V(\varphi)\) be approximated by a polygonal
piecewise linear approximation, as shown in FIG.~\ref{fig:VPhi}.

Let us first develop the idea for the single field case, dropping the field index $i$ and introducing the segment
index for the field values \(\tilde \varphi_s, s = 1, \ldots, N\), such that the two minima reside at \(\tilde \varphi_{1, N}\).
The values of the potential are \(\tilde V_s = V(\tilde \varphi_s)\) and the linear segments are
\begin{align} \label{eqVlin}
  V_s(\varphi) &= \underbrace{\left(\frac{\tilde V_{s+1} - \tilde V_s}{\tilde \varphi_{s+1} - \tilde \varphi_s}\right)}_{8 \, a_s}
  \left( \varphi - \tilde \varphi_s \right) + \tilde V_s  - \tilde V_N.
\end{align}
For linear \(V_s\), the exact solution of~\eqref{eqBounceEqD} on the section \(s\) is
\begin{align} \label{eqPBSol}
  \quad \varphi_s(\rho) &= v_s + \frac{4}{D} a_s \rho^2 + \frac{2}{D-2} \frac{b_s}{\rho^{D-2}},
\end{align}
with \(D > 2\). Two dimensions require minor modifications derived
in~\ref{appD268}. Because we are dealing with a finite number of segments, the solution either
\renewcommand{\labelenumi}{\alph{enumi})}
\begin{enumerate}
  \item starts from \(\varphi_0\) at \(\rho = 0\) with \(\dot \varphi_1 = 0\), which gives
  %
  \begin{align} \label{eqInitCondDCaseA}
    \qquad v_1 &= \varphi_0, & b_1 &= 0,
  \end{align}
  %
  \item or waits at \(\tilde \varphi_1\) until \(\rho = R_0\), which translates into
  \begin{align} \label{eqInitCondDCaseB}
    v_1 &= \tilde \varphi_1 - \frac{4}{D-2} a_1 R_0^2, & b_1 &= \frac{4}{D} a_1 R_0^D.
  \end{align}
\end{enumerate}
Regardless of the initial condition, the field in the final section \(\varphi_{N-1}\) stops in the second minimum
\(\tilde \varphi_N\) at some final radius \(R_{N-1}\) such that
\begin{align} \label{eqbvFinalDg2}
  v_{N-1} &= \tilde \varphi_N - \frac{4}{D - 2} a_{N-1} R_{N-1}^2,
  \\
  b_{N-1} &= \frac{4}{D} a_{N-1} R_{N-1}^D,
\end{align}
where \(a_0 = a_N = 0, \) because the first derivatives are zero in the minima. Thus there is no issue with
the \(\rho \to 0\) limit: in case a) the singularity of the friction term is regulated by \(b_1 = 0\), while in the case b)
there is no singularity to start with and \(R_0\) is non-zero. Similarly, the role of \(\rho \to \infty\) is taken over
by the final radius \(R_{N-1}\) that is finite and numerically under control, see~\ref{appRootFReR}
for details.

The total Euclidean action of the bounce is then a sum of linear parts
\begin{equation} \label{eqSEFinalDg2}
  S_D = {\mathcal T}_D + {\mathcal V}_D,
\end{equation}
with the integrated kinetic and potential pieces
\begin{align} \label{eqTFinalDg2}
  \begin{split}
    {\mathcal T}_{>2} &=\frac{2 \pi^{\frac{D}{2}}}{\Gamma \left(\frac{D}{2}\right)}
    \sum_{s=0}^{N-1} \Biggl[\rho^2 \biggl(\frac{32 a_s^2 \rho^D}{D^2(D+2)} - \frac{8}{D} a_s b_s -\\
    &\frac{2 b_s^2}{\rho^D (D-2)} \biggr)\Biggr]_{R_{s-1}}^{R_s},
    \end{split}
        \\
    \label{eqVFinalDg2}
    \begin{split}
    {\mathcal V}_{>2} &= \frac{2 \pi^{\frac{D}{2}}}{\Gamma \left(\frac{D}{2}\right)}
     \sum_{s=0}^{N-1} \Biggl[ \frac{\rho^D}{D} \biggl( 8 a_s \left(v_s - \tilde \varphi_s \right) +
      \\
    &
    \tilde V_s -\tilde V_N \biggr) + \rho^2 \biggl(\frac{32 a_s^2  \rho^D}{D(D+2)} + \frac{8 a_s b_s}{D-2}\biggr)
    \Biggr]_{R_{s-1}}^{R_s},
     \end{split}
\end{align}
which is valid for both instances, a) and b), with the understanding that \(R_{-1} = 0\) and in case a) \(R_0 = 0\).
To determine the action above the field segments \(\varphi_s(\rho)\) of the bounce need to be computed. To this end, a
segmentation of \(\{\tilde \varphi_s \}\) is set up, such that given the \(V(\tilde \varphi_s)\), the \(a_s\) parameters
are fixed by~\eqref{eqVlin}. We shall return to the choice of segmentation procedure in \S\ref{subsecImplement} below.
What remains to be calculated are the \(v_s\), \(b_s\) and the unknown radii \(R_s, s = 0,\ldots, N-1\).

We now demonstrate that solving the PB is a single variable problem, i.e. once the initial radius is known,
the entire solution is determined. The free parameters are fixed by matching conditions required to glue neighbouring
linear bounces into a single smooth solution, as in FIG.~\ref{fig:VPhi}. There are three conditions, two for the field value
\(\varphi_s(R_s) = \tilde \varphi_{s+1} = \varphi_{s+1}(R_s)\) to match onto the initial segmentation at \(R_s\) and
another one for the derivative \(\dot \varphi_s(R_s) = \dot \varphi_{s+1}(R_s)\)
\begin{align} \label{eqMatchField1Dg2}
  v_s + \frac{4}{D} a_s R_s^2 + \frac{2}{D-2} \frac{b_s}{R_s^{D-2}} &= \tilde \varphi_{s+1},
  \\ \label{eqMatchField2Dg2}
  v_{s+1} + \frac{4}{D} a_{s+1} R_s^2 + \frac{2}{D-2} \frac{b_{s+1}}{R_s^{D-2}} &= \tilde \varphi_{s+1},
  \\ \label{eqMatchDerivDg2}
  \frac{4}{D} \left(a_{s+1} - a_s \right) R_s^D + b_s - b_{s+1} &= 0.
\end{align}
These three conditions per segment precisely determine the unknown \(v_s, b_s \) and \(R_s\). Therefore, one can
increase the number of sections at will without introducing additional free parameters.

The two-segment \(N=3\) problem can be solved analytically in some instances, as shown in~\ref{appN3D}.
With more segmentation points, one can transform the system into a single variable problem that can be solved
numerically. For some particular \(D\), further simplifications are possible.

Let us derive the recursion relations for \(R_s(v_s, b_s)\) such that they can be computed numerically. We first
derive \(v_s\) and \(b_s\) by subtracting~\eqref{eqMatchField1Dg2} from~\eqref{eqMatchField2Dg2} and using~\eqref{eqMatchDerivDg2}
\begin{align} \label{eqvbnDg2}
  v_s &= v_1 - \frac{4}{D-2} \sum_{\sigma=1}^{s-1} \left(a_{\sigma+1} - a_\sigma \right) R_\sigma^2,\\
  b_s &= b_1 + \frac{4}{D} \sum_{\sigma = 1}^{s-1} \left(a_{\sigma+1} - a_\sigma \right) R_\sigma^D.
\end{align}
The individual radii can be solved directly from~\eqref{eqMatchField1Dg2}
\begin{align} \label{eqRnDg2}
  a_s R_s^D - \frac{D}{4} \delta_s R_s^{D-2} +
  \frac{D}{2(D - 2)} b_s = 0,
\end{align}
with \(\delta_s = \tilde \varphi_{s+1} - v_s\). Resulting Eq.~\eqref{eqRnDg2} is a fewnomial with simple
closed form solutions
\begin{align} \label{eqRnD3}
 & D = 3: & \quad  2 R_s  &= \frac{1}{\sqrt{a_s}} \left( \frac{\delta_s}{\xi} + \xi\right),
 \\
  &  & \xi^3 & = \sqrt{36 a_s b_s^2 - \delta_s^3} - 6  \sqrt{a_s} b_s,
  \\ \label{eqRnD4}
 & D = 4: & \quad  2 R_s^2 &= \frac{1}{a_s} \left(\delta_s + \sqrt{\delta_s^2 - 4 a_s b_s}\right).
\end{align}
The radii corresponding to \(D = 2, 6, 8\) can be found in Eqs~\eqref{eqRnD2}-\eqref{eqRnD8}
of~\ref{appD268}. This concludes the analytical setup of the PB construction.

\paragraph{Derrick's theorem for piecewise actions.} A well known result due to Derrick~\cite{Derrick:1964ww}
is the relation between the integrated kinetic and potential parts in~\eqref{eqTFinalDg2} and~\eqref{eqVFinalDg2}.

We will use this theorem to find the PB solution and to test the goodness of the approximation,
so let us recall its essential point. For the action to remain minimal upon rescaling the argument of the solution to
\(\varphi(\rho/\lambda)\), the following identity has to hold
\begin{align} \label{eqSDerrickCont}
  S_D^{(\lambda)} &= \lambda^{D-2} \mathcal T + \lambda^D \mathcal V,
  \frac{d S_{D}^{(\lambda)}}{d \lambda} \biggr|_{\lambda = 1} = 0  \Rightarrow \\
  &(D-2) \mathcal T + D \mathcal V = 0.
\end{align}

For piecewise actions, such as the PB under consideration, the above identity is
modified because~\eqref{eqSEFinalDg2} becomes a sum of finite integration intervals.
While rescaling \(\rho \to \rho/\lambda\) has no effect on integration limits in the continuous limit,
rescaling the finite intervals \(R_s \to R_s/\lambda \) in~\eqref{eqSEFinalDg2} introduces a
manifest \(\lambda\) dependence. As a result,
\begin{align} \label{eqSDerrickPiece}
  S_{D, PB}^{(\lambda)} &= \sum_s \left( \lambda^{D-2} \mathcal T^{(\lambda)}_s +
   \lambda^D \mathcal V^{(\lambda)}_s \right),\\
   \mathcal T^{(\lambda)}_s &\propto \int_{\frac{R_{s-1}}{\lambda}}^{\frac{R_s}{\lambda}}
   \rho^{D-1} {\rm d} \rho \, \dot \varphi_s^2,
\end{align}
and similarly for \(\mathcal V^{(\lambda)}_s\). Imposing the vanishing derivative of the polygonal
\(S_{D, PB}^{(\lambda)}\) over \(\lambda\), one obtains a complicated finite version of the identity in~\eqref{eqSDerrickCont},
modifying the relation between \(\mathcal T^{(\lambda)}_s\) and \(\mathcal V^{(\lambda)}_s\).
However, with a sufficiently large number of segments, the relation~\eqref{eqSDerrickCont}
with \(\mathcal T \to \sum_s \mathcal T^{(\lambda)}_s\) and \(\mathcal V \to \sum_s \mathcal V^{(\lambda)}_s\)
is quickly recovered.

At the same time, one can use the continuous version of~\eqref{eqSDerrickCont} with the
input potential (not the polygonal approximation) to verify the goodness of the polygonal
solution. This is shown on the right side of FIG.~\ref{fig:RfDk} in~\ref{appRootFReR},
where about a permille level is achieved with \(N=400\) segments.

%
%
\section{Evaluating polygonal bounces} \label{secEvaluate}

\subsection{Implementation} \label{subsecImplement}

\paragraph{Overview.} Let us turn to the implementation of the PB method. In the work of~\cite{Duncan:1992ai},
the bounce equations were cast into an algebraic system and solved in a closed form. The approach followed 
here instead is to recursively compute the bounce parameters and solve a single boundary condition equation.

The boundary equation is obtained by combining~\eqref{eqbvFinalDg2} with~\eqref{eqvbnDg2}
and setting \(s = N-1\), which leads to
\begin{align} \label{eqMatchbN}
  \sum_{\sigma = 0}^{N-1} \left(a_{\sigma+1} - a_\sigma \right) R_\sigma^D &= 0,
\end{align}
valid for all \(D\). Because the \(R_s\) are already solved for, the final condition for \(v_{N-1}\) holds automatically.
Alternatively, one can use the relation in~\eqref{eqSDerrickCont} with the polygonal potential, and look
for the solution of
\begin{align} \label{eqLambda}
  \lambda = \sqrt{\frac{(2-D)\mathcal{T}}{D \mathcal{V}}} = 1.
\end{align}
In order to solve the boundary equation, either~\eqref{eqMatchbN} or~\eqref{eqLambda}, one has to find the
initial radius \(R_s\) from which the subsequent \(v_s, b_s, R_s\) are computed recursively until
the boundary condition is satisfied. This is the algebraic analog of the shooting method used to
solve~\eqref{eqBounceEqD} directly.

Adding more segmentation points improves the accuracy of the approximation, but does not
exponentially increase the computational burden, timing scales linearly with \(N\).

\paragraph{Segmentation.}
To set up the polygonal potential approximation, one chooses a set of field values \(\{ \tilde \varphi_s \}\) 
that interpolate between the positions between which the tunneling happens, as exemplified in FIG.~\ref{fig:VPhi}.
Throughout this work we assume the original potential \(V(\varphi)\) to be non-pathological in the
sense that it admits at least one bounce solution between these two values\footnote{The polygonal
approach can also be applied to unbounded potentials with a local minimum at \(\tilde \varphi_N\). In
such instance, case b) does not exist, since the field cannot wait at the true minimum. Instead, the choice
of the exit point, i.e. \(\tilde \varphi_1\) must be deep enough for the field, starting from \(\varphi_0\),
to roll down to the false minimum.}.

To describe an arbitrary potential, enough segments should be taken to capture all the non-linearities
with desired precision. In addition, the action converges faster if the segmentation is tailored to a specific
potential, i.e. if the density of points increases close to the extrema. This geometrical insight is a particular 
feature of the polygonal approach and allows for intuitive understanding of the problem prior to the actual 
calculation of the bounce.

\begin{figure*}[!ht]
  \centering
  \includegraphics[width=1.75\columnwidth]{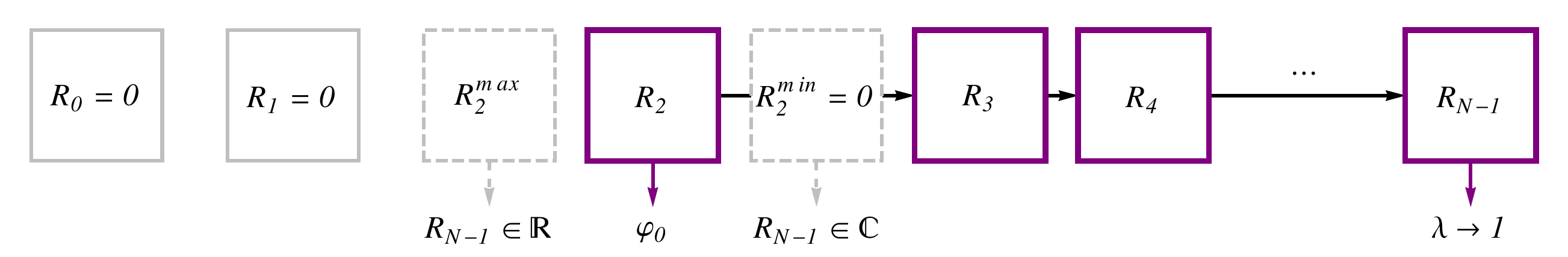}
  \caption{Schematic overview of finding the PB. The segment with the solution (in
  this example \(s = 2\) and \(R_{in} = R_2\)) can be found by evaluating the PB on the boundaries of
  \(R_2^{\min} = 0\) and \(R_2^{\max}\) and checking that the imaginary part of the final radius \(R_{N-1}\)
  becomes non-zero. Finally, the solution of \(R_2\) is found such that the scaling parameter \(\lambda \to 1\).}
  \label{fig:Scheme}
\end{figure*}

For a sufficiently large \(N\) the specific choice of coverage is not relevant, the na\"ive uniform distribution 
reproduces any reasonable potential when \( N \to \infty \) and converges smoothly to the final value.
In this limit, the resolution of \( \Delta \tilde \varphi_s \) is small enough such that \(\varphi_0\) always falls 
above \(\tilde \varphi_1\) and only case a) persists. This is to be expected because such limit is equivalent 
to the original problem in~\eqref{eqBounceEqD} where \(R_0 \to 0\) and only \(\varphi_0\) matters.

\paragraph{Computing the initial bounce radius.}

With a given segmentation at hand one has to find the initial radius \(R_{in}\) that solves the boundary equation.
Actually, the task can be simplified by a priori isolating the field segment on which the solution exists.

One can see from the right panel of FIG.~\ref{fig:VPhi} that the list of Euclidean radii \(\{R_s\}\), must be real,
positive and growing (the true minimum is on the left by convention). On the other hand, Eq.~\eqref{eqMatchbN}
contains a number of nested roots and becomes progressively non-linear as \(N\) grows and generically admits
complex solutions for the radii.

Let us demonstrate that the final radius \(R_{N-1}\) becomes imaginary as \(R_{in}\) is varied across the
true solution. This can be understood by noticing that the discriminant \(\delta_{N-1}^2 - 4 a_{N-1} b_{N-1}\)
in~\eqref{eqRnD4} vanishes due to the boundary conditions in~\eqref{eqbvFinalDg2}, likewise for \(D=3\).
Thus, when one expands the discriminant around the true solution, only the linear term
remains, which will flip the sign of the discriminant and thereby the imaginary part of the final radius appears,
as seen on the left panel of FIG.~\ref{fig:RfDk} and shown schematically on FIG.~\ref{fig:Scheme}.

Furthermore, note that in both cases a) and b) one only needs to solve for \(R_{in}\), from which the initial field 
value \(\varphi_0\) can be determined. In case b) this is merely the position of the minimum \(\varphi_0 = \tilde \varphi_1\), 
while in case a), it is obtained from \(R_{in}\) and~\eqref{eqMatchField1Dg2}
\begin{equation} \label{eqRin}
  \varphi_0 = \tilde \varphi_{in+1} - \frac{4}{D} a_{in} R_{in}^2.
\end{equation}
From here one can infer the interval for \(R_{in} \in [0, R_{in}^{\max}]\) by setting \(\varphi_0\) to the lower and upper 
boundary of the segment in~\eqref{eqRin}. The way to find the segment with the solution a priori is therefore to evaluate 
the final radius from these two limiting \(R_{in}\) and checking whether it becomes imaginary, as illustrated in FIG.~\ref{fig:Scheme}.

Once the segment containing the solution has been found, one can proceed to solve the polygonal
bounce by solving either~\eqref{eqMatchbN} or~\eqref{eqLambda}. Another approach is to take advantage
of the fact that the bounce solution depends solely on \(R_{in}\). This is a dimensional parameter, which can
therefore be rescaled by the optimal amount computed from~\eqref{eqLambda}, which essentially aims to
minimize the action. For example, one may begin with \(R_{in}^{\max}\), compute the corresponding \(\lambda\),
which in general will be different from 1, and proceed by iteration from \(R_{in} = \lambda R_{in}^{\max}\).
This procedure converges in a few iterations to a permille level. Alternatively, one can solve~\eqref{eqLambda} 
with standard root finding algorithms.

By increasing the number of segments, the initial radius (e.g. \(R_0\) in case b)) decreases until \(R_{in} = 0\),
when the domain of the solution disappears and one has to switch to the next segment. This agrees
with~\eqref{eqBounceEqD}, as does the fact that the final radius \(R_{N-1}\) grows steadily to infinity
when \(N \to \infty\), see FIG.~\ref{fig:Phi0Rfs}.

%
%
\subsection{Examples, convergence and comparisons} \label{subsecExamples}

\paragraph{Linearly displaced quadratic potential} \hspace{-1em} is the benchmark potential to
test the PB method. It is defined as in the work of Coleman~\cite{Coleman:1977py}
\begin{equation} \label{eqVC}
  V(\varphi) = \frac{\lambda}{8} \left(\varphi^2 - v^2 \right)^2 + \varepsilon \left(\frac{\varphi - v}{2 v} \right),
\end{equation}
and shown on the left panel of FIG.~\ref{fig:VPhi}. For convenient numerical evaluation, we set \(\lambda = 0.25, v = 1\); 
other points in parameter space can be obtained by rescaling~\cite{Sarid:1998sn}. For such choice 
of parameters, varying \(\varepsilon\) from 0.01 to 0.08 covers all the regions of interest, starting from thin wall regime of 
small \(\varepsilon\), going to well separated minima until the second minimum disappears.

\begin{figure}
  \includegraphics[width=.8 \columnwidth]{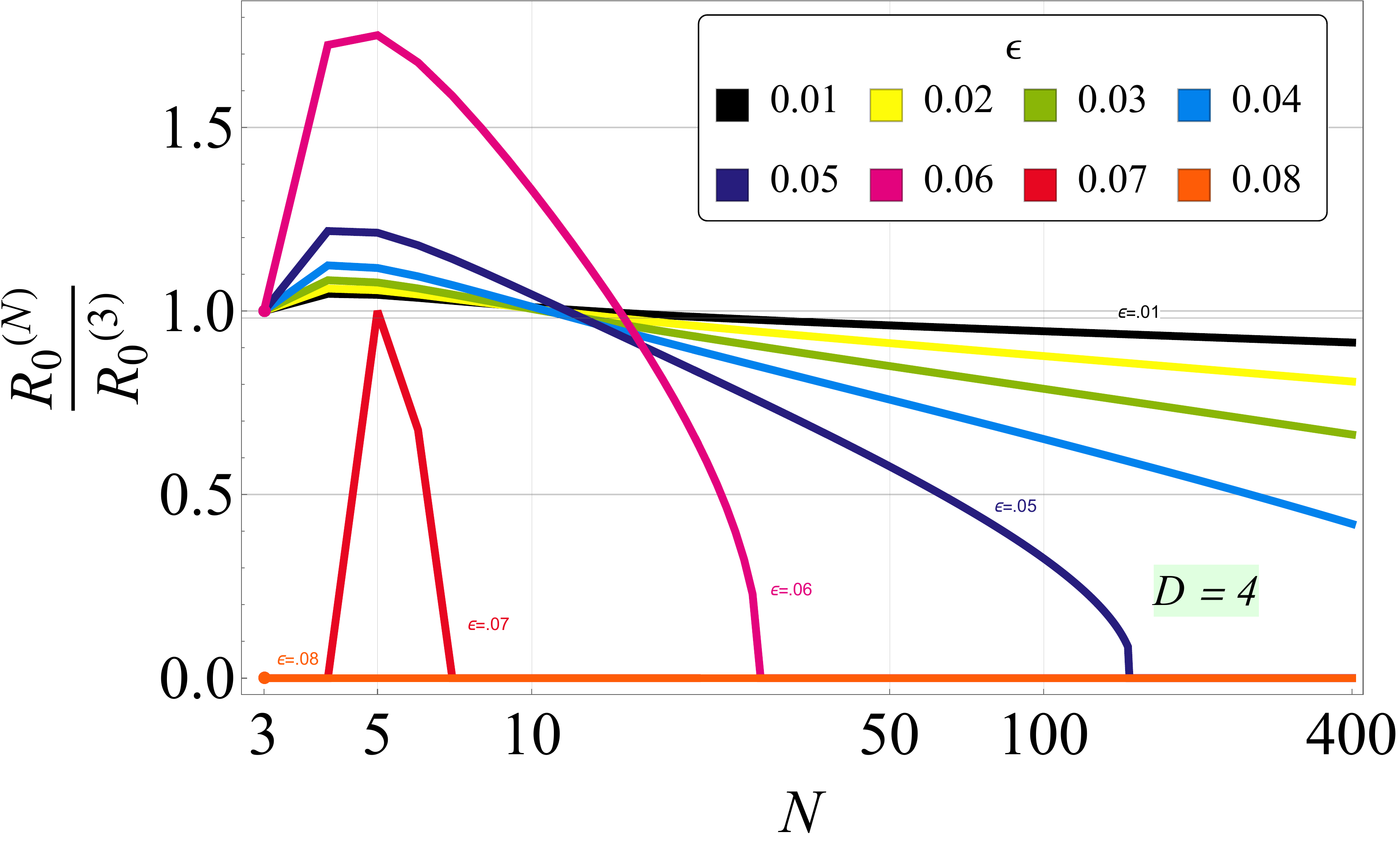}
  \caption{The initial radius \(R_0^{(N)}\) of case b) and \(D=4\) for the uniform segmentation with \(N\) points,
  normalized to the minimal \(N=3\) setup. Similar behavior appears for \(D=3\).
  Different lines correspond to the range of \(\varepsilon\) which controls the separation between the minima
   in~\eqref{eqVC}, see text for details.}
  \label{fig:Rs}
\end{figure}

We now apply the PB method to the potential in~\eqref{eqVC}, employing the homogeneous
segmentation for simplicity. The first results are the \(\varphi_0\) and \(R_0\) that attempt to
solve~\eqref{eqMatchbN}. The solution for \(R_0\) varies with \(N\), therefore we show the behavior
of \(R_0^{(N)}/R_0^{(3)}\) in FIG.~\ref{fig:Rs}, where \(R_0^{(N)}\) is the initial radius corresponding to
some fixed \(N\). For any choice of \(\varepsilon\), the \(R_0\) decreases with \(N\) and eventually drops to
zero, as seen in FIG.~\ref{fig:Rs}. At this point, one has to switch from b) to a)\footnote{This is true in general
when \(N\) is sufficiently large. The reverse transition from a) to b) is also possible when \(N\) is small enough
and a particular segmentation is chosen. This happens for \(\varepsilon = 0.07\) in \(D = 4\) shown
on the right panel of FIG.~\ref{fig:Rs}.}.

The smaller \(\varepsilon\) is, the closer one goes towards the thin wall regime, where the field needs
to wait close to the minimum. This means \(R_0\) remains sizeable for higher values of \(N\)
and one needs to introduce many segments for \(R_0\) to reach zero, as clear from FIG.~\ref{fig:Rs}.
On the other hand, the transition from b) to a) happens faster when \(\varepsilon\) increases.
Finally, when \(\varepsilon\) is large enough, the transition eventually disappears and we are left with
case a) right from the start at \(N=3\).

The number of dimensions also has an impact on the transition from b) to a), as seen in FIG.~\ref{fig:Rs}.
Keeping \(\varepsilon\) fixed, the transition in \(D=4\) occurs for higher \(N\) with respect to \(D=3\).
This is expected because the damping term in~\eqref{eqBounceEqD} is proportional
to \(D\) and thus becomes more important in higher dimensions.

\begin{figure}
  \includegraphics[width = 1 \columnwidth]{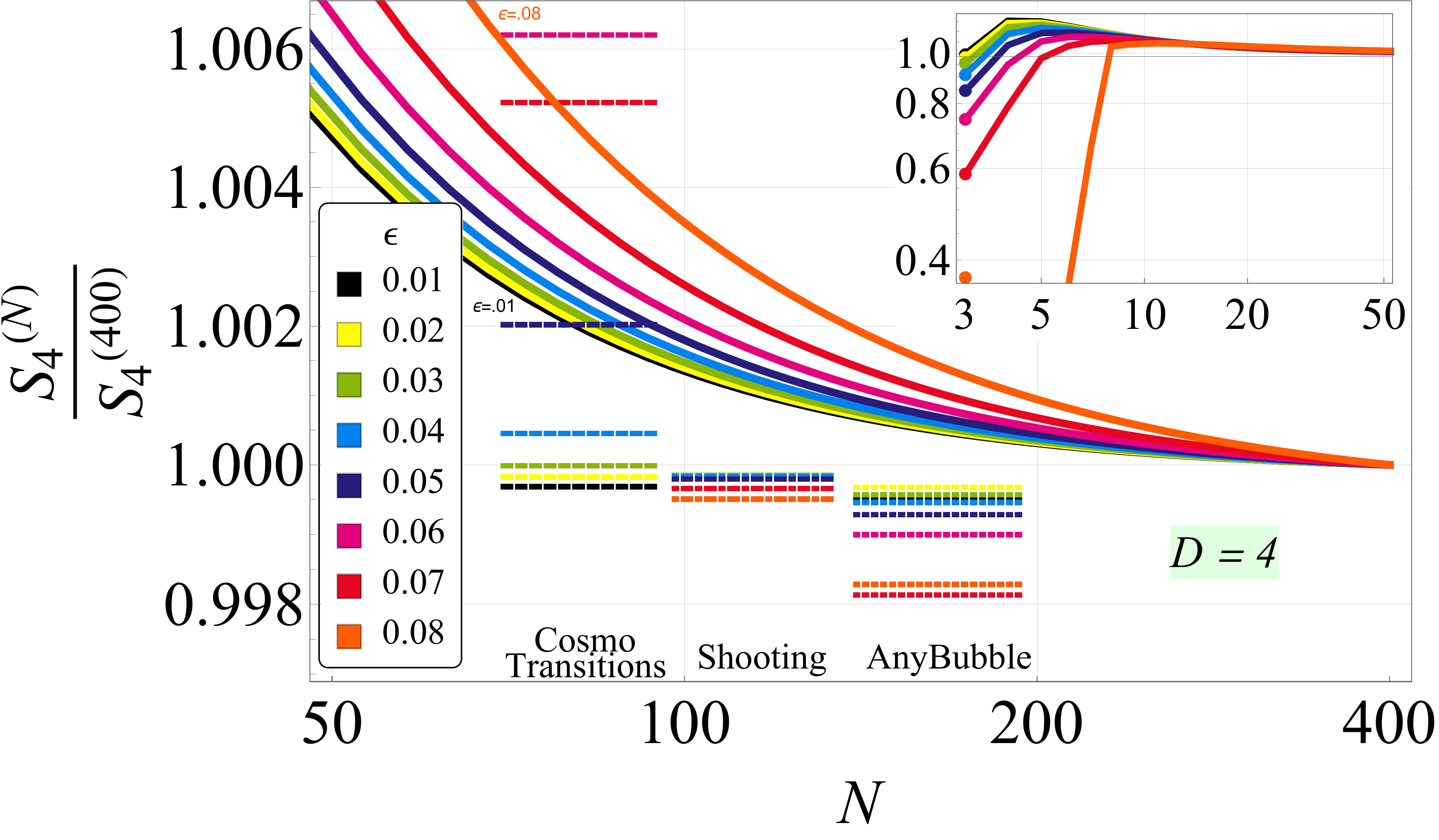}
  \caption{The bounce action \(S_D^{(N)}\) normalized to the maximal \(N=400\) uniform segmentation with \(D=4\). 
  The solid lines show the PB method for different \(\varepsilon\) that defines the input potential. The inset shows the same,
   for a smaller number of segments. The dotted lines show the comparison to other methods and tools, see text for details.}
  \label{fig:Ss}
\end{figure}

The final step after obtaining \(R_0\) or \(\varphi_0\) is to compute the main object of interest: the Euclidean action
\(S_D\) in~\eqref{eqSEFinalDg2} that sets the bubble nucleation rate. FIG.~\ref{fig:Ss} shows the main point of 
this work: the convergence of \(S_D^{(N)}\), the action for \(N\) segments with \(D=4\) (the results are basically 
the same for \(D=3\)). The \(S_D^{(N)}\) is normalized to the large \(N=400\) value in order to ease the comparison 
between different \(\varepsilon\).

In the limit of \(\varepsilon \simeq 0\) one ends up in the thin wall regime, and therefore \(N=3\)
has to produce the correct result of~\cite{Duncan:1992ai}, in agreement with the inset of FIG.~\ref{fig:Ss}.
With increasing \(\varepsilon\), the potential in~\eqref{eqVC} will eventually lose the second minimum. For
any potential close to this threshold, the resolution of the homogeneous segmentation has to be precise enough
to describe the local maximum, otherwise the solution cannot exist a priori.
This is precisely what happens in FIG.~\ref{fig:Ss} for \(\varepsilon = 0.08\), the \(N = 4\) segmentation
is too rough to possess an intermediate maximum.
In general, the approximation worsens for \(4 \leq N < \mathcal O(10)\), which is an artefact of the assumed uniform
segmentation.  Conversely, for higher \(N\), the action starts to converge rapidly and the rate is faster in case b) for smaller
\(\epsilon\), where the shooting method instead becomes increasingly unstable.

The initial approximations with small \(N\)s, shown in the inset, are already quite close to the end result,
and are valid at about 10\% level. It is clear that the \(N=3\) segmentation always {\em underestimates}
the action and this simple approximation becomes progressively better as \(\varepsilon\) decreases. On the
other hand, as \(N\) increases, the method starts to {\em overestimate} the bounce and converges to the final
result from above. Even for moderate \(N=10\) the accuracy of the estimation is below 10\% and goes below
the permille level when \(N=200\). The convergence is slightly faster for \(N=3\), moreover the rate
of convergence can be improved by choosing an appropriate segmentation.

To compare the PB method to existing methods, we show the results of other
approaches in FIG.~\ref{fig:Ss}. The other three calculations are the usual shooting method of Eq.~\eqref{eqBounceEqD} and
the out-of-the-box results from \textsc{CosmoTransitions}~\cite{Wainwright:2011kj} and
\textsc{AnyBubble}~\cite{Masoumi:2016wot} packages. Note that in these examples all the methods
agree within a few permille level.

\begin{figure} [!h]
  \includegraphics[width=.46\columnwidth]{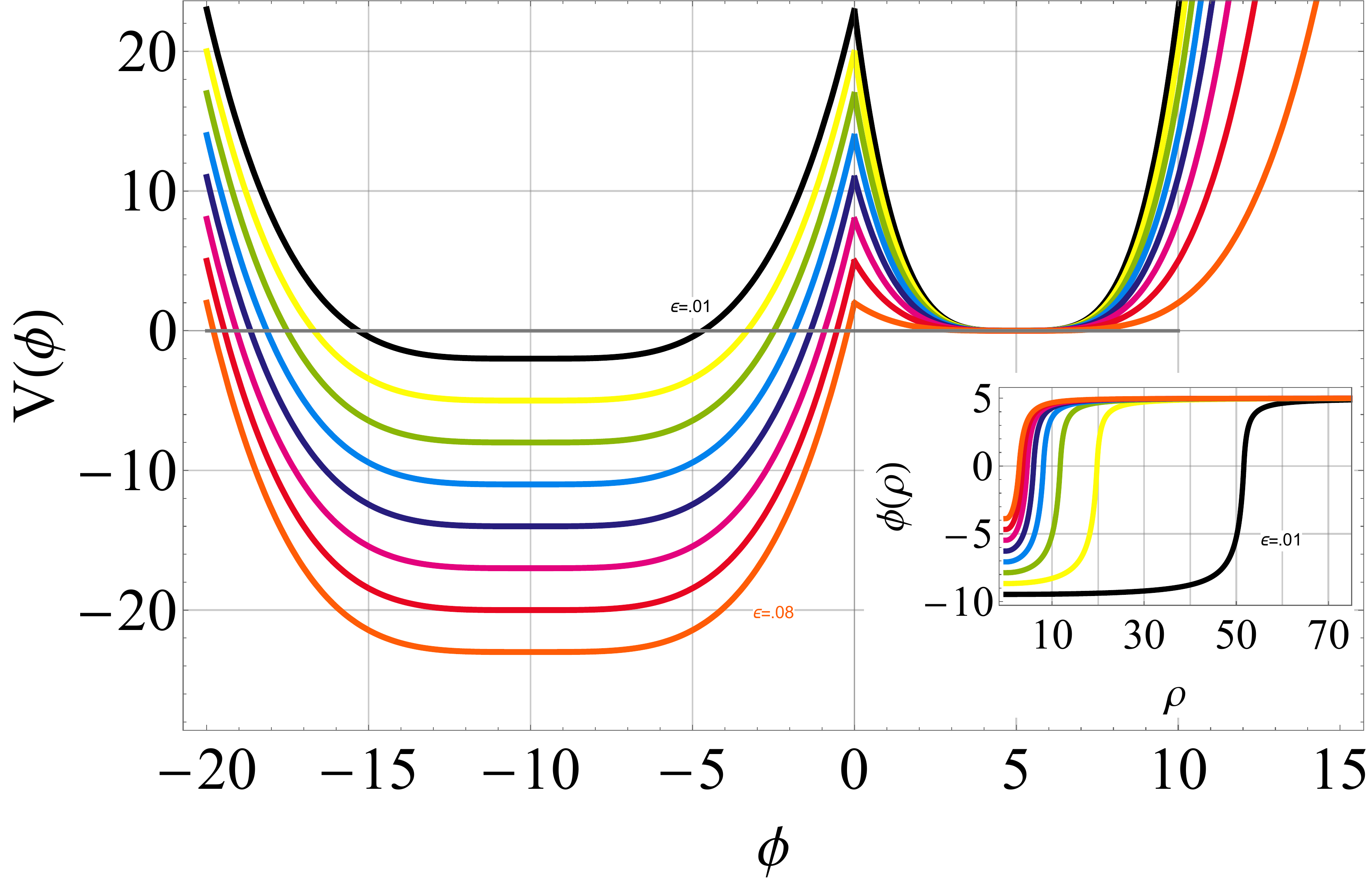} %
  \includegraphics[width=.515\columnwidth]{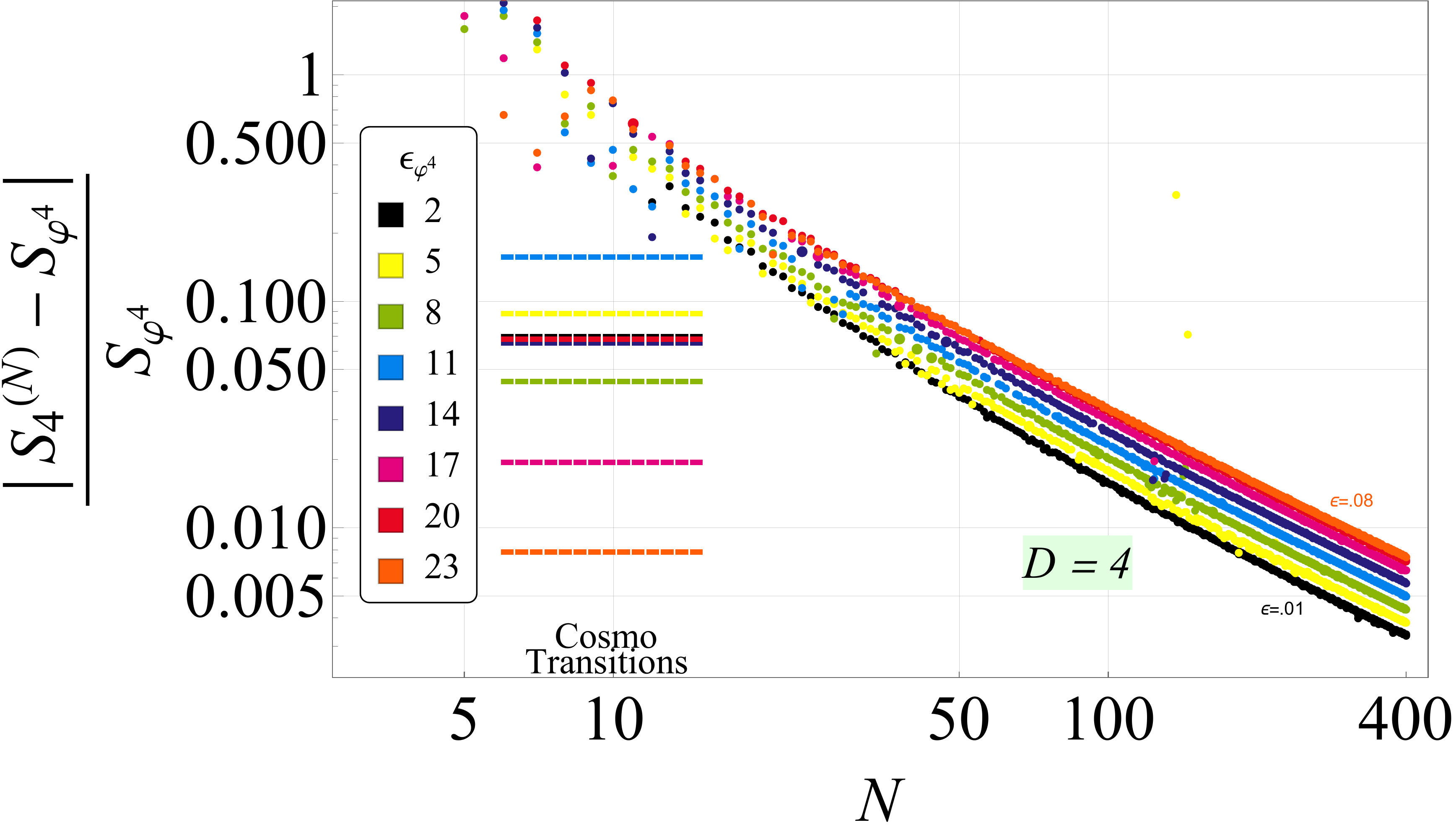}
  \caption{Left: The bi-quartic potential with the bounce field solution in the inset. Right:
  The PB action normalized to the exact value (see text for details).}
  \label{figBiQuart}
\end{figure}

%
%
\paragraph{Bi-quartic potential} \hspace{-1em} is another example of a simple but non-trivial exact
solution~\cite{Dutta:2011rc} that glues two quartic field functions. Since the bi-quartic bounce is computed
analytically, it can serve as an additional test of the polygonal method.

The bi-quartic potential is parameterized by \(\varepsilon_{\varphi^4}\) representing the gap in the potential
difference, which varies from the thin to thick wall regime, as shown in FIG.~\ref{figBiQuart}.
The presence of the cusp creates issues for standard approaches based on the shooting method,
due to the non-differentiable potential. On the other hand, the polygonal method turns out to be quite robust 
and the solution can always be found. Nevertheless, for smooth convergence it is convenient to employ a 
bi-uniform segmentation on both sides of the cusp.

The resulting bounce action is shown on the right of FIG.~\ref{figBiQuart} and goes below percent level
accuracy with \(\mathcal O(100)\) field segments. The dashed lines also provide the comparison to
\textsc{CosmoTransitions}\footnote{We were unable to recover the value of the action by using \textsc{CosmoTransitions}
out-of-the-box. Instead, we extrapolated the field bounce solution to manually compute the action shown on
FIG.~\ref{figBiQuart}. Still, this procedure failed for lower values of \(\varepsilon_{\varphi^4}\) closer to the thin
wall regime. Moreover, computing the bounce using \textsc{AnyBubble} was not possible for any value of \(\varepsilon_{\varphi^4}\).}.

\paragraph{Other potentials and additional minima.}
%
We also tested the PB approach on the potential in~\eqref{eqVC}, corrected with the logarithmic field
dependence. With such a deformed potential, the calculation proceeds exactly as before and the polygonal
approximation works as expected.

The method was successfully applied on examples with further intermediate local minima.
Such a situation may arise when more fields are involved and a particular path in field space is chosen.
The prototype of \(N=5\) with two triangles and a single additional minimum can be considered
as the minimal setup illustrating such situations. As long as the bounce exists, i.e. if the intermediate
minimum is not too deep,~\eqref{eqMatchbN} gives a consistent real solution.

%
%
\section{Quantum fluctuations with polygonal bounces} \label{secFluctuate}
The simplicity of the semi-classical polygonal solution can be exploited also for computing the quantum
corrections, i.e. the prefactor of the decay rate, originally derived in~\cite{Callan:1977pt}. A number of
studies on the prefactor proposed different numerical methods in $D=3$~\cite{Strumia:1998nf, Munster:1999hr}
and $D=4$~\cite{Baacke:2003uw, Dunne:2005rt}, and recent progress has been made on precision
calculations in presence of gauge interactions~ \cite{Andreassen:2016cvx}, scale invariant instantons
and extended gauge theories~\cite{Chigusa:2018uuj}. On the other hand, not many explicit analytical
results on the prefactor are available with a notable exception of the thin wall limit~\cite{Konoplich:1986zp}.

The total decay rate at one loop is
\begin{align}
  \Gamma = \left(\frac{S_4}{2 \pi} \right)^2 \left|\frac{\det'(-\partial^2 + V''(\varphi(\rho)) )}{
  \det(-\partial^2 + V''(\varphi_-))} \right|^{-1/2} \, e^{-S_4 - \delta_4},
\end{align}
where $S_4$ is the semi-classical action computed from the bounce solution $\varphi(\rho)$
and $\det'$ is the determinant of the fluctuation operator, i.e. the product of its eigenvalues, with
zeros removed. Finally, $\delta_4$ is the perturbative one loop counterterm of the action that
absorbs the renormalization infinities.

In computing the determinant, we follow the work of Dunne~\cite{Dunne:2005rt}, where the
fluctuation operator $\mathcal O$, i.e. the second variation of the action, is decomposed in a multipole
expansion due to the $O(4)$ symmetry
\begin{align}
  \mathcal O_l &= - \frac{d^2}{d \rho^2} - \frac{3}{\rho}\frac{d}{d \rho}  + \frac{l(l+1)}{\rho^2}+ V''(\rho) + 1,
  \\
  V''(\rho) &= - 3 \varphi(\rho) + \frac{3 \alpha}{2} \varphi^2(\rho),
\end{align}
where $V''(\rho)$ is the rewritten form of~\eqref{eqVC} with the removed asymptotic value of 1, such
that for the fluctuations around the true vacuum $ V''(\tilde \varphi_1) = 0$.

Instead of computing all the eigenfunctions $\psi_l(\rho)$ of $\mathcal O_l$ and summing the corresponding
eigenvalues, it is convenient to use the Gel'fand-Yaglom theorem~\cite{Gelfand:1959nq} that relates the ratio of
determinants to the value of the ratio of eigenfunctions evaluated at infinite Euclidean time
\begin{align}
  \frac{\det \mathcal O_l}{\det \mathcal O_l^{\text{free}}} &= \mathcal R_l(\rho = \infty)^{(l+1)^2},
  &
  \mathcal R_l(\rho) &= \frac{\psi_l(\rho)}{\psi_l^{\text{free}}(\rho)}.
\end{align}

The calculation of the pre-factor splits in two parts: the low $l$ region up to an arbitrary $l \leq L \simeq \mathcal O(10)$
and the high $l$ region, going to infinity. In the low $l$ part the ratio of determinants $\mathcal R_l$ is computed by
solving the partial differential equation for $\mathcal R_l$, because the solutions $\psi_l^{\text{free}}$ for the fluctuations around
the true vacuum are known Bessel functions $\psi_l^{\text{free}} = I_{l+1}(\rho)/\rho$.

The bounce solution $\varphi(\rho)$ determines the shape of $V''(\rho)$, and in the low $l$ regime, the contribution to the decay rate is
finite and proportional to the sum of the log of all the ratios of determinants:
\begin{align}
  - \ln \Gamma_{\text{lo}} = \frac{1}{2} \sum_{l=0}^{L} \left( l + 1 \right)^2 \ln \left| \mathcal R_l(\infty) \right|.
\end{align}

On the other hand when $l > L \gg 1$ one can solve for the $\mathcal R_l$ using the WKB approximation~\cite{Dunne:2005rt},
which is regularized with the proper counter terms in $\delta_4$ (with optional higher orders~\cite{Hur:2008yg} for
faster convergence). This high-$l$ part of the rate, i.e. $- \ln \Gamma_{\text{hi}}$ is
\begin{align}
   -\frac{(L + 1) (L + 2)}{8} \mathcal I_1 + \frac{\ln L}{16} \mathcal I_2 -
  \frac{\mathcal I_2 + \mathcal I_3}{16} ,
\end{align}
with the three relevant integrals given by
\begin{align}
  \mathcal I_1 &= \int_0^\infty {\rm d} \rho \, \rho \, V'' \left(\rho \right),
  \\
  \mathcal I_2 &= \int_0^\infty {\rm d} \rho \, \rho^3 \, V'' \left( V'' + 2 \right),
  \\
  \mathcal I_3 &= \int_0^\infty {\rm d} \rho \, \rho^3 \, V'' \left(V'' + 2 \right) \ln \frac{\rho}{2}.
\end{align}
These integrals are straightforward to compute analytically and the total prefactor contribution is the sum of the low and high $l$ pieces.

\begin{figure}[!h]
  \centering
  \includegraphics[width=.485\columnwidth]{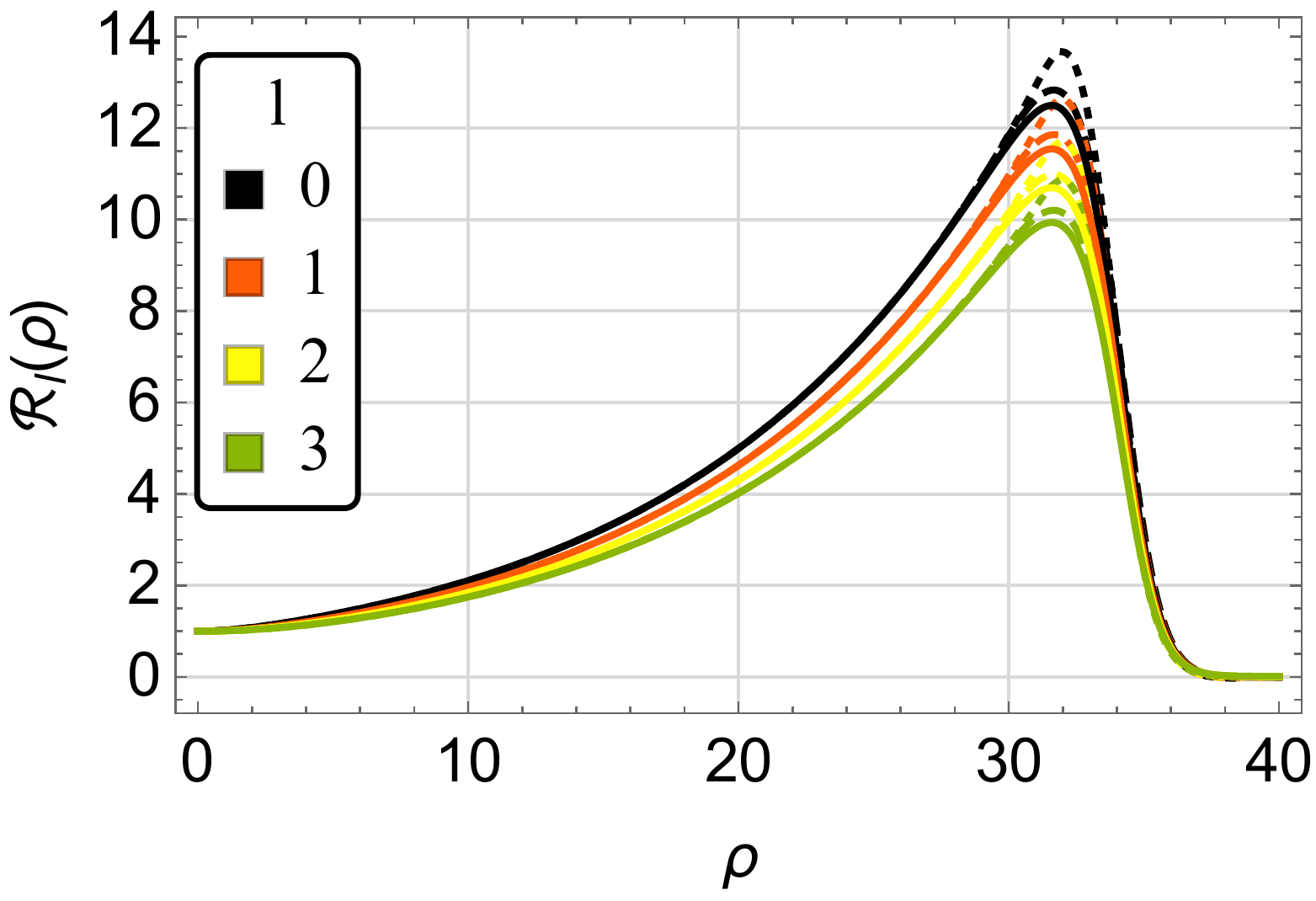} %
  \includegraphics[width=.49\columnwidth]{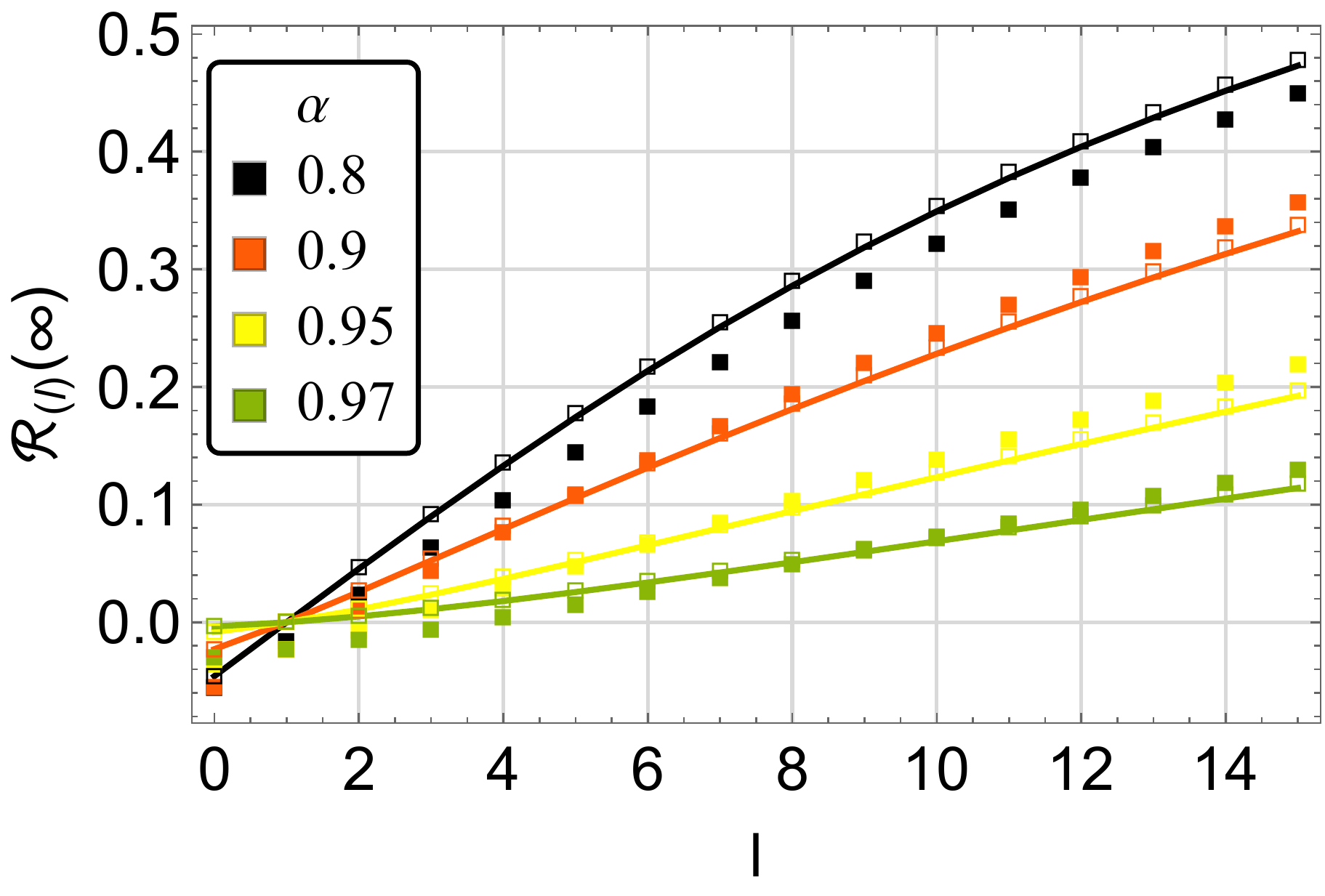}
  \caption{The ratio of determinants $\mathcal R_l$ for a given multipole. Left: The $\rho$ dependence
  for shooting in solid and the PB solution with $N=3$ ($N=50$) in dotted (dashed) lines.
  Right: The ratio at $\rho \to \infty$ with solid (empty) squares denoting the $N=3$ ($N=50$) PB
  approximation, while the solid line connects the results from the shooting procedure.}
  \label{figRlsN3}
\end{figure}

\begin{table}
  \begin{tabular}{r | c | c | c | c | c}
  \hline \hline
  $\alpha$  & shooting & $N=3$ & $N=10$ & $N=50$ & $N=100$
  \\ \hline
  0.8 		& 0.36	&  0.30 & 0.31 & 0.31 & 0.30
  \\
  0.9 		& 0.30 	& 0.24 & 0.27 & 0.27 & 0.28
  \\
  0.95 	& 0.24 	& 0.20 & 0.22 & 0.23 & 0.23
  \\
  0.97 	& 0.22 	& 0.18 & 0.20 & 0.21 & 0.21
  \\
  \hline \hline
  \end{tabular}
  \label{tabTotGam}
  \caption{The total prefactor contribution at one-loop, computed using the numerical shooting procedure and compared
  with the polygonal method with $N=3, 10, 50$ and 100 segmentation points. The rate is normalized to $(1-\alpha)^3$
  and agrees with the analytical thin wall limit result~\cite{Konoplich:1986zp} that gives $9/32 (1 - 2 \pi/(9 \sqrt 3 )) \sim 0.17$.}
\end{table}

The crucial component in computing $\mathcal R_l$ is of course the semi-classical bounce solution. In
FIG.~\ref{figRlsN3} we show the resulting $\mathcal R_l$ using the precise numerical shooting solutions and the
PB approximation with the minimal $N=3$ and the more precise $N=50$.

The values at infinity $\mathcal R_l(\infty)$ agree with the expectation of a single negative eigenvalue
for $l =0$, four-fold degenerate zero for $l=1$ and the rest of $l \geq 2$ being positive. This is true for the precise
shooting procedure, however the $N=3$ PB bounce produces a number of negative eigenvalues, while for
$N=50$ the correct spectrum is recovered. This happens because the semi-classical solution is not approximating
the exact potential with sufficient precision, the proof for one negative and multiple zero eigenvalues~\cite{Coleman:1987rm}
(and the entire calculation of the fluctuations) relies on the fact that the semi-classical action is extremised.
Nevertheless, blithely summing the absolute values of $\mathcal R_l(\infty)$ gives a rather precise (and very simple)
estimate of the decay rate prefactor, as seen in TAB.~\ref{tabTotGam}.

The crude $N=3$ approximation fails when $\alpha \ll 1$, however it works well in the thin wall limit when $\alpha \to 1$
and all of the approaches coincide, as shown on the right panel of FIG.~\ref{figRlsN3}.

%
%
\section{Extending polygonal bounces} \label{secExtend}

\begin{figure}[!h]
  \centering
  \includegraphics[width=.5\columnwidth]{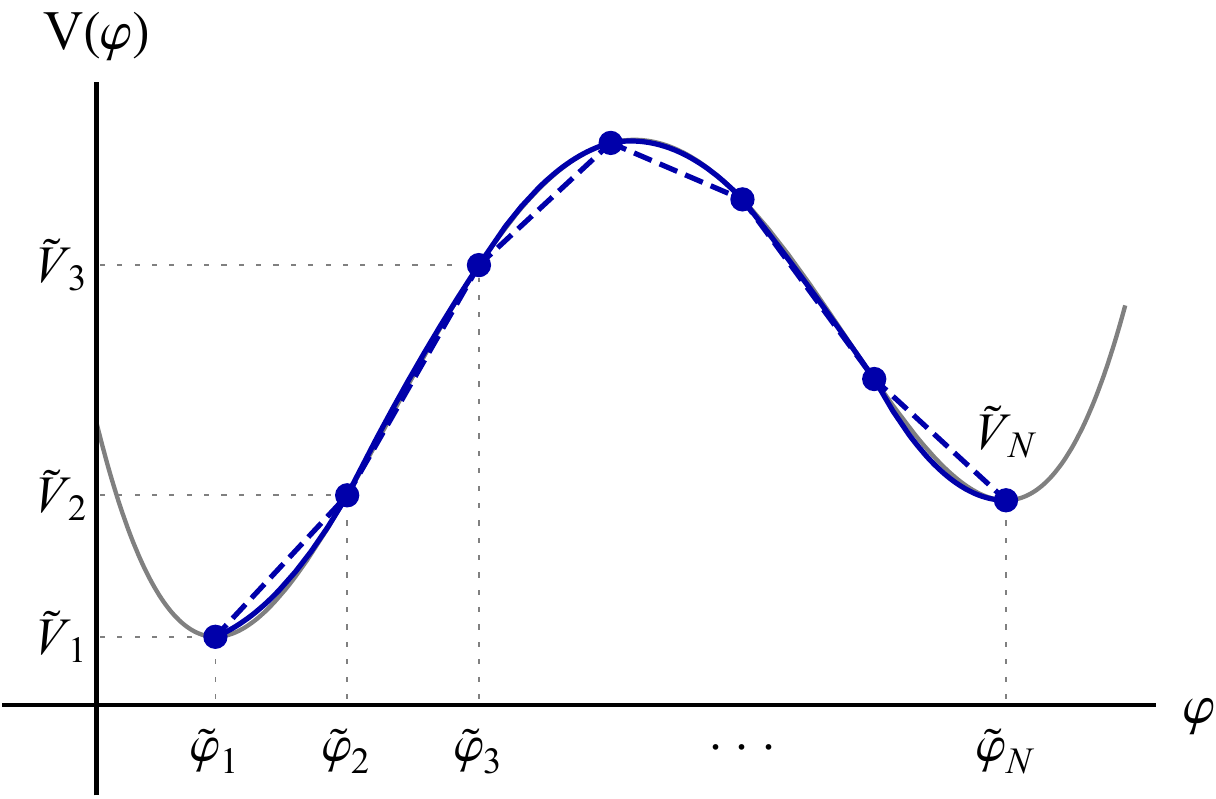}%
  \includegraphics[width=.5\columnwidth]{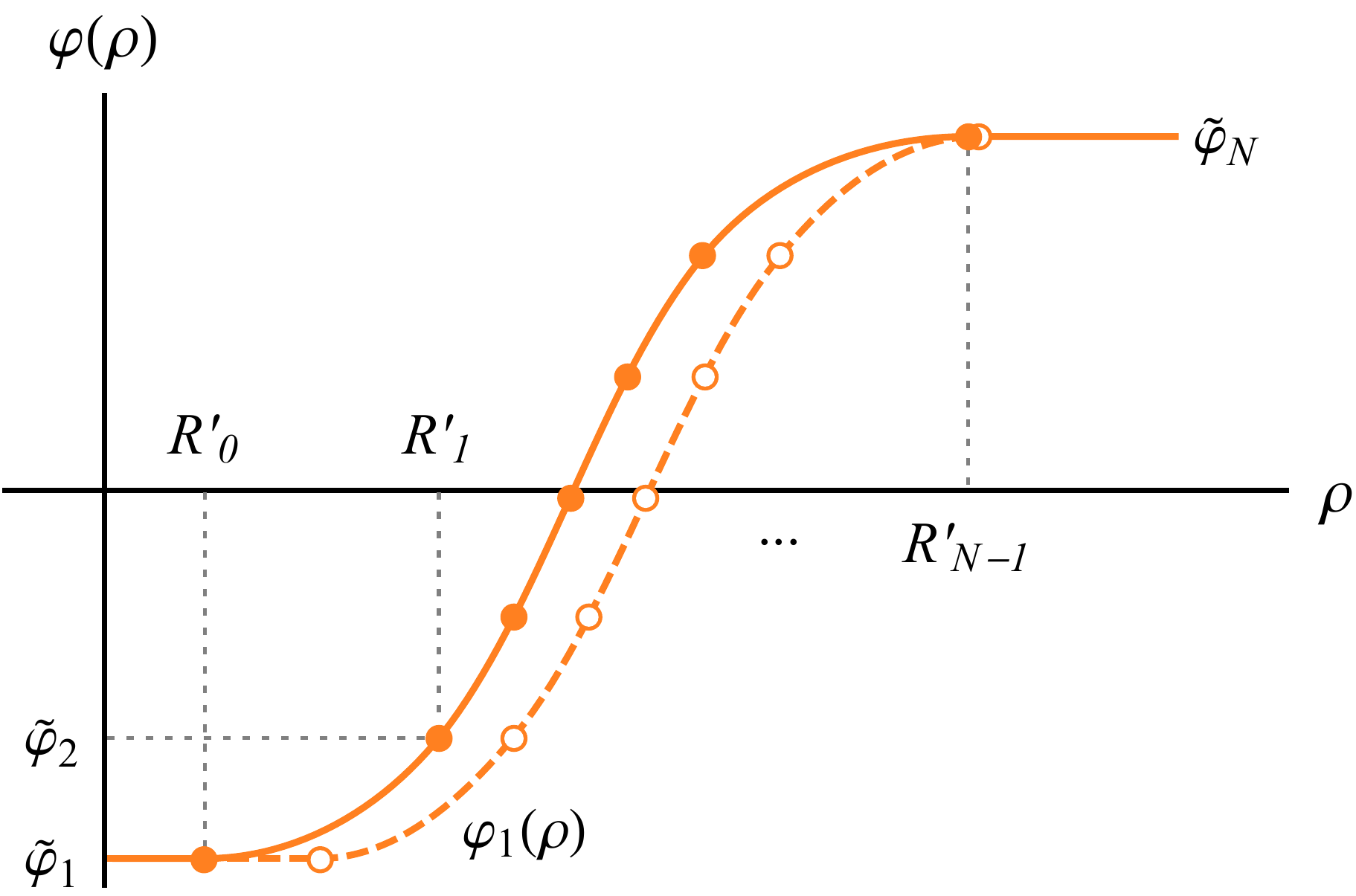}
  \caption{
    Left: The linearly off-set quartic potential in gray, the linear polygonal approximation with \(N=7\) in
    dashed blue and the \(2^{\text{nd}}\) order quadratic correction in solid blue.
    Right: The field solution in the PB approximation in dashed and the $2^{\text{nd}}$ order improved
    solution in solid orange.}
  \label{figNLOBounce}
\end{figure}

In this section we develop a general procedure of including non-linear corrections to the PB.
This is done by setting up a systematic procedure based on the Taylor expansion of the potential and
then building the new bounce solution perturbatively on the PB ansatz.

Higher order corrections describe non-linear features that are not there in
the leading approximation, for example around the extrema of \(V\) where the linear
part of the potential vanishes. Although the PB solution is formally exact when \(N \to \infty\), the nonlinear
corrections may enhance the convergence of the action, depending on the type of the
potential and the order to which we correct.

\paragraph{Generalities.} Consider the complete bounce solution expanded around the
PB: \(\varphi = \varphi_{PB} + \xi\), such that the correction to the potential is evaluated on the PB
background and the bounce equation becomes
\begin{align}
  \ddot \varphi + \frac{D-1}{\rho} \dot \varphi &= 8 \left(a + \alpha \right) + \delta dV\left(\varphi_{PB}(\rho)\right),
  \\
  \ddot \xi + \frac{D-1}{\rho} \dot \xi &= 8 \alpha + \delta dV(\rho), 
  \\
  \delta dV &= dV(\varphi_{PB}(\rho)) - 8 \left( a + \alpha \right),
\end{align}
where \(\alpha\) is an arbitrary linear part. The bounce correction \(\xi\) is then given by
\begin{align}
 & \xi = \nu + \frac{2}{D-2} \frac{\beta}{\rho^{D-2}} + \frac{4}{D} \alpha \rho^2 + \mathcal I(\rho),\\
 & \mathcal I(\rho)= \int_{\rho_0}^\rho \text{d}y \, y^{1-D} \int_{\rho_1}^y \text{d} x  \, x^{D-1} \delta d V(x).
\end{align}
Evaluating the above integral \(\mathcal I\) for an arbitrary \(\delta d V\) and computing the unknown parameters
of \(\xi\) is involved and basically equivalent to the numerical integration of~\eqref{eqBounceEqD}.
However, a systematic expansion of the potential and linearization simplify this approach considerably.

%
%
\paragraph{Perturbation.} On a given segment, the potential can be expanded in Taylor series
around $\tilde \varphi_s$
\begin{equation}\label{eqVTaylor}
  \tilde V_s -  \tilde V_N + \partial \tilde V_s \left(\varphi_s - \tilde \varphi_s \right) +
  \frac{\partial^2 \tilde V_s}{2} \left(\varphi_s - \tilde \varphi_s \right)^2 + \ldots,
\end{equation}
where the constants \(\partial V_s, \partial^2 V_s, \ldots\) are determined by matching the values and (higher)
derivatives of \(V\). When \(N\) increases, the segmentation becomes arbitrarily dense and thus the terms
beyond the linear one in~\eqref{eqVTaylor} become progressively negligible.

To illustrate this point, we expand \(V\) to second order
\begin{align} \label{eqaalpha1}
  \partial \tilde V_s &= 8 \left(a_s + \alpha_s\right), \quad 
  8 \alpha_s = 8 a_s - d \tilde V_{s+1},
  \\
  \partial^2 \tilde V_s &= \frac{d \tilde V_{s+1} - \partial \tilde V_s}{\tilde \varphi_{s+1} - \tilde \varphi_{s}} =
  \frac{d \tilde V_{s+1} - 8\left(a_s + \alpha_s\right)}{\tilde \varphi_{s+1} - \tilde \varphi_{s}},
\end{align}
where \(d \tilde V_s \) stands for the derivative of the original potential evaluated at \(\tilde \varphi_s\).
This is the additional information required from the original potential in order to get to the next-to-leading
order. The \(\alpha_s\) coefficients are thereby fixed and the inclusion of the quadratic correction improves the
fit of the potential near the extrema, as seen from FIG.~\ref{figNLOBounce}. Moreover, with a large \(N\),
one has \(\alpha_s \ll a_s\) as clear from~\eqref{eqaalpha1}, which is consistent
with the assumption of perturbativity.

With this approximation of the potential, the non-homogeneous part of the correction is
\begin{align}
  \mathcal I_s &= \int_{\rho_0}^\rho \text{d}y \, y^{1-D} \int_{\rho_1}^y \text{d} x
  \, x^{D-1}  \partial^2 \tilde V_s \left(\varphi_{PB s} - \tilde \varphi_s \right),
\end{align}
which can be evaluated for \(D = 3, 4\)
\begin{align}
  \mathcal I_s^{D=3} &= \partial^2 \tilde V_s \left(\frac{v_s - \tilde \varphi_s}{6} \rho^2 +
  b_s \rho + \frac{a_s}{15} \rho^4 \right),
  \\
  \mathcal I_s^{D=4} &= \partial^2 \tilde V_s \left(\frac{v_s - \tilde \varphi_s}{8} \rho^2 +
  \frac{b_s}{2} \ln \rho + \frac{a_s}{24} \rho^4 \right),
\end{align}
where the arbitrary integration constants \(\rho_0, \rho_1\) were chosen to simplify the expression
for \(\mathcal I_s\) without loss of generality because they can be absorbed in \(\nu_s, \beta_s\).
The remaining task is to compute the unknown coefficients \(\nu_s, \beta_s\) and the new matching radii by
requiring the solution to be continuous and differentiable as in the PB case.

Given that \(\varphi_{PB}\) and its matching radii are already close to the actual solution,
the new radii have to be close to the previous ones
\begin{align}
  R_s &\to R_s\left(1 + r_s \right), & r_s \ll 1.
\end{align}
Following the same procedure as in the PB construction above, we set up the modified initial, final
and matching conditions for the correction \(\xi\).
These conditions are then perturbatively linearized in \(r_s\) to get the recursion relations
for the parameters
\begin{align} \label{eqnui}
  \begin{split}
  \nu_s & = \nu_1 - \sum_{\sigma=1}^{s-1} \left(\frac{2}{D-2} \frac{\beta_{\sigma+1} - \beta_\sigma}{R_\sigma^{D-2}} \right. +
  \\
  & \left. \frac{4}{D} \left(\alpha_{\sigma+1} - \alpha_\sigma \right) R_\sigma^2 + \mathcal I_{\sigma+1} - \mathcal I_\sigma \right),
  \end{split}
  \\
  \label{eqbetai}
  \begin{split}
  \beta_s & = \beta_1 + \sum_{\sigma=1}^{s-1} \left(\frac{4}{D} \left(\alpha_{\sigma+1} - \alpha_\sigma \right) + \right.
  \\
  & \left. 4 r_\sigma \left(a_{\sigma+1} - a_\sigma \right) + \frac{\dot {\mathcal I}_{\sigma+1} \dot {\mathcal I}_\sigma }{2 R_\sigma} \right) R_\sigma^D,
  \end{split}
\end{align}
and similarly a linear equation for the radius correction at each segment is
\begin{equation}
  r_s = \frac{\beta_s + \frac{D-2}{2}\left(\nu_s + \mathcal I_s + \frac{4}{D} \alpha_s R_s^2 \right)R_s^{D-2}
  }{(D-2) \left(b_s - \frac{4}{D} a_s R_s^D \right)}.
\end{equation}

\begin{figure}
  \includegraphics[width = .9\columnwidth]{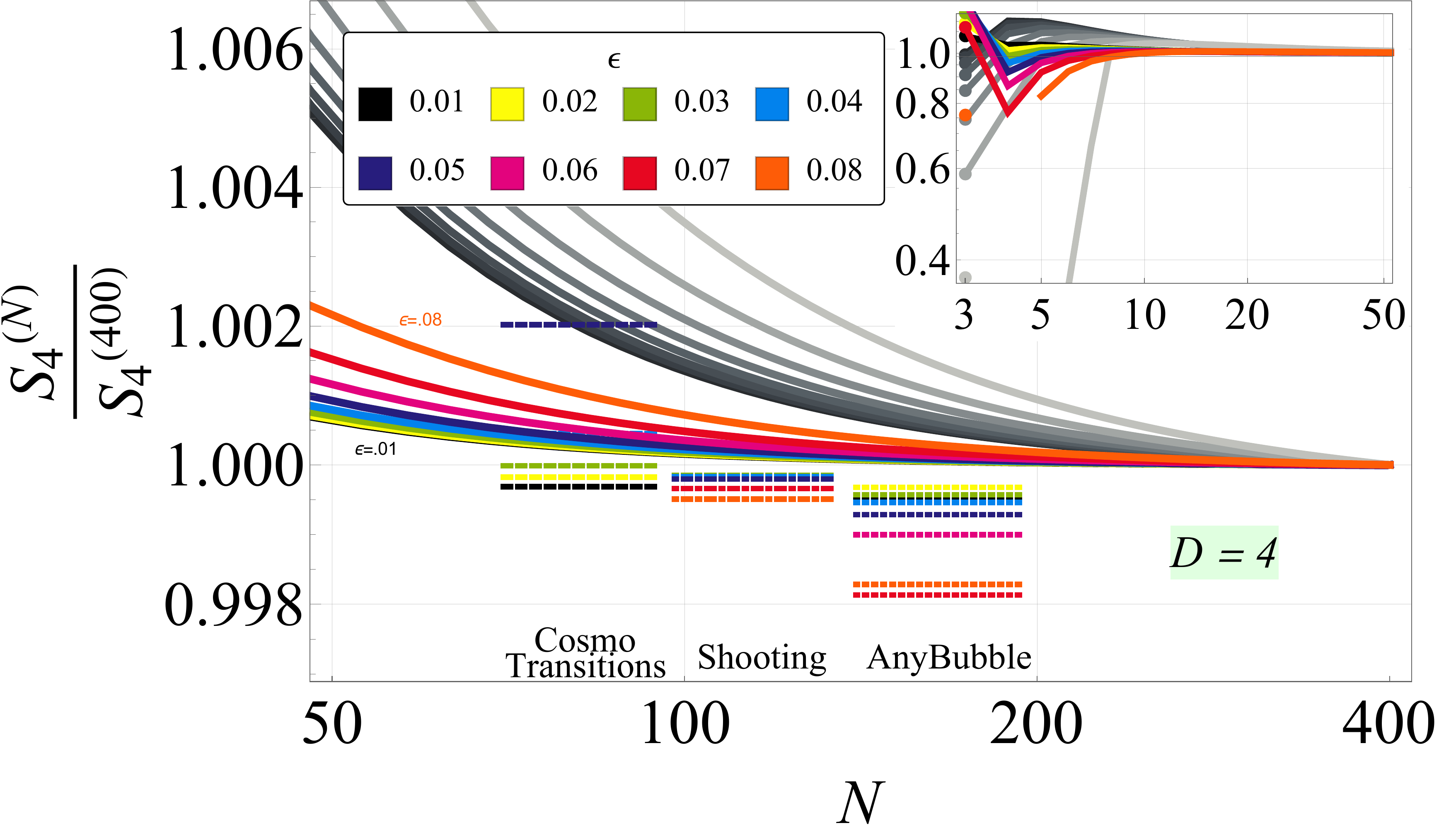} %
  \caption{ The bounce action of the improved bounce calculation including the second order correction. The lower 
  colored lines correspond to the corrected action, while the upper gray ones show the leading PB for comparison.}
  \label{fig:SNLO}
\end{figure}

Following the same logic as in the PB case above, we compute the initial radius correction \(r_{in}\)
by solving the linear equation that satisfies the final matching condition. Being a linear equation,
this additional step does not require significant computing time but improves the accuracy
of the action and speeds up convergence.

\paragraph{Improved action.} To understand the effect of second order corrections,
we reconsider the usual displaced quartic potential and show the improved action in FIG.~\ref{fig:SNLO}.
The correction significantly improves the approximation of the action by nearly an order
of magnitude improvement for any given \(N\) and \(\varepsilon\). In other words, to achieve the same level
of accuracy one needs to consider half as many segments.

Because the polygonal bounce perturbation requires only to solve a linear equation, the computational cost
of computing the bounce solution with a given accuracy is reduced significantly. Moreover, the final result
of the bounce field configuration is again given in the form of segmented analytical functions, which allows
for further manipulation.

%
%
\section{Multi-field polygonal bounces} \label{secMultiField}
%
%
Computing the false vacuum decay rate with multiple scalar fields faces a number of technical difficulties.
These are related to the fact that the Euclidean action is not a minimum but a saddle point. In terms of the
bounce solution, one has to look for the (fine-tuned) initial condition in the higher dimensional field space
and then integrate the coupled system of differential equations, usually numerically.

Existing approaches to this problem~\cite{Kusenko:1995jv, John:1998ip, Cline:1999wi, Wainwright:2011kj,
Konstandin:2006nd, Park:2010rh, Akula:2016gpl, Masoumi:2016wot, Espinosa:2018szu, Athron:2019nbd}
address these challenges in various ways. In general, solutions where the shooting and path deformation are
decoupled exhibit oscillatory (and therefore slower) path convergence, multifield shooting face non-linear
scaling with the number fields, and most approaches have difficulties with thin wall regimes and provide
purely numerical output of the bounce field configuration, as well as the Euclidean action.

\begin{figure*} \hspace{1cm}
  \includegraphics[width = .8\columnwidth]{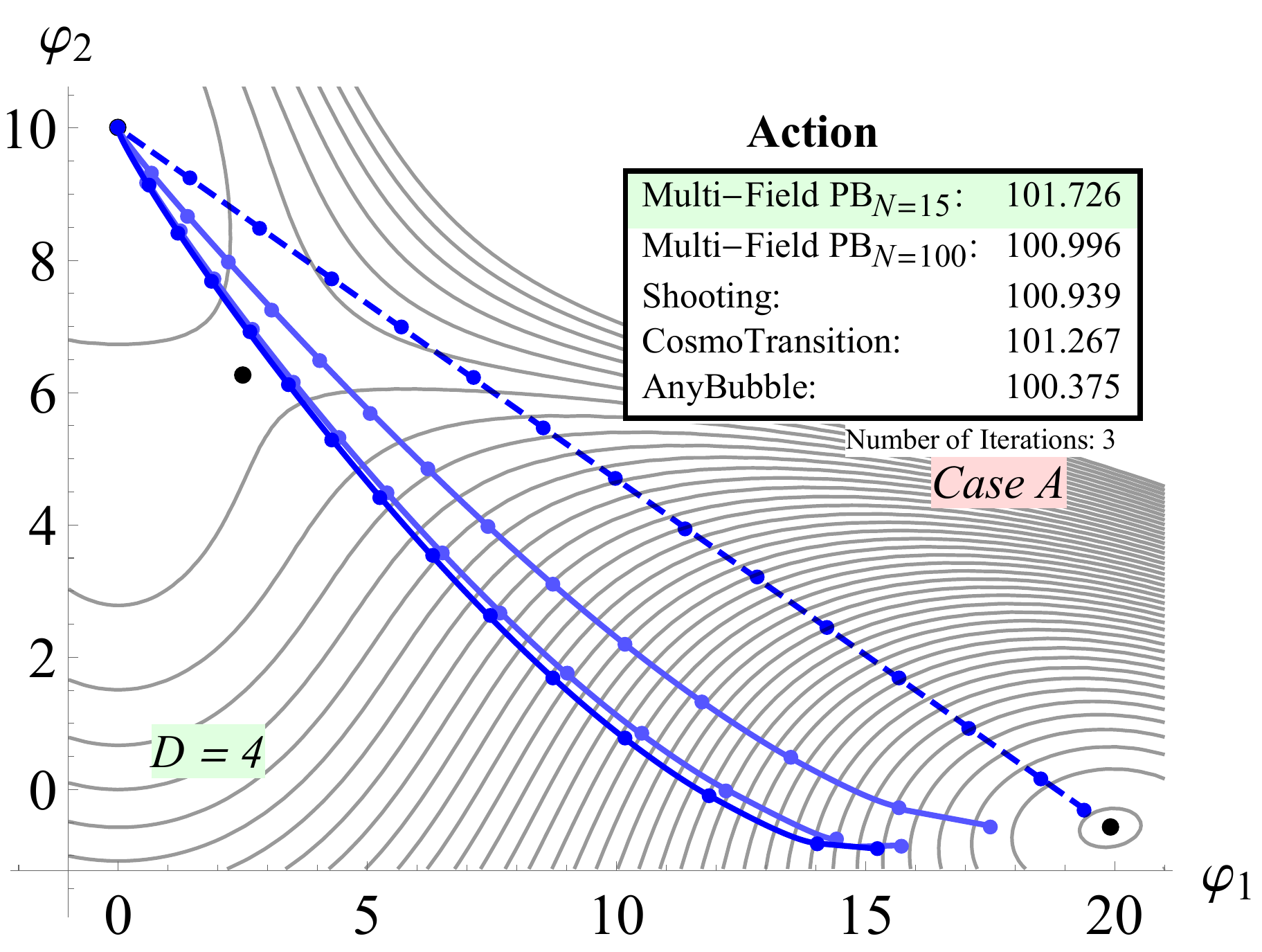}
  \hfill
  \includegraphics[width = .8\columnwidth]{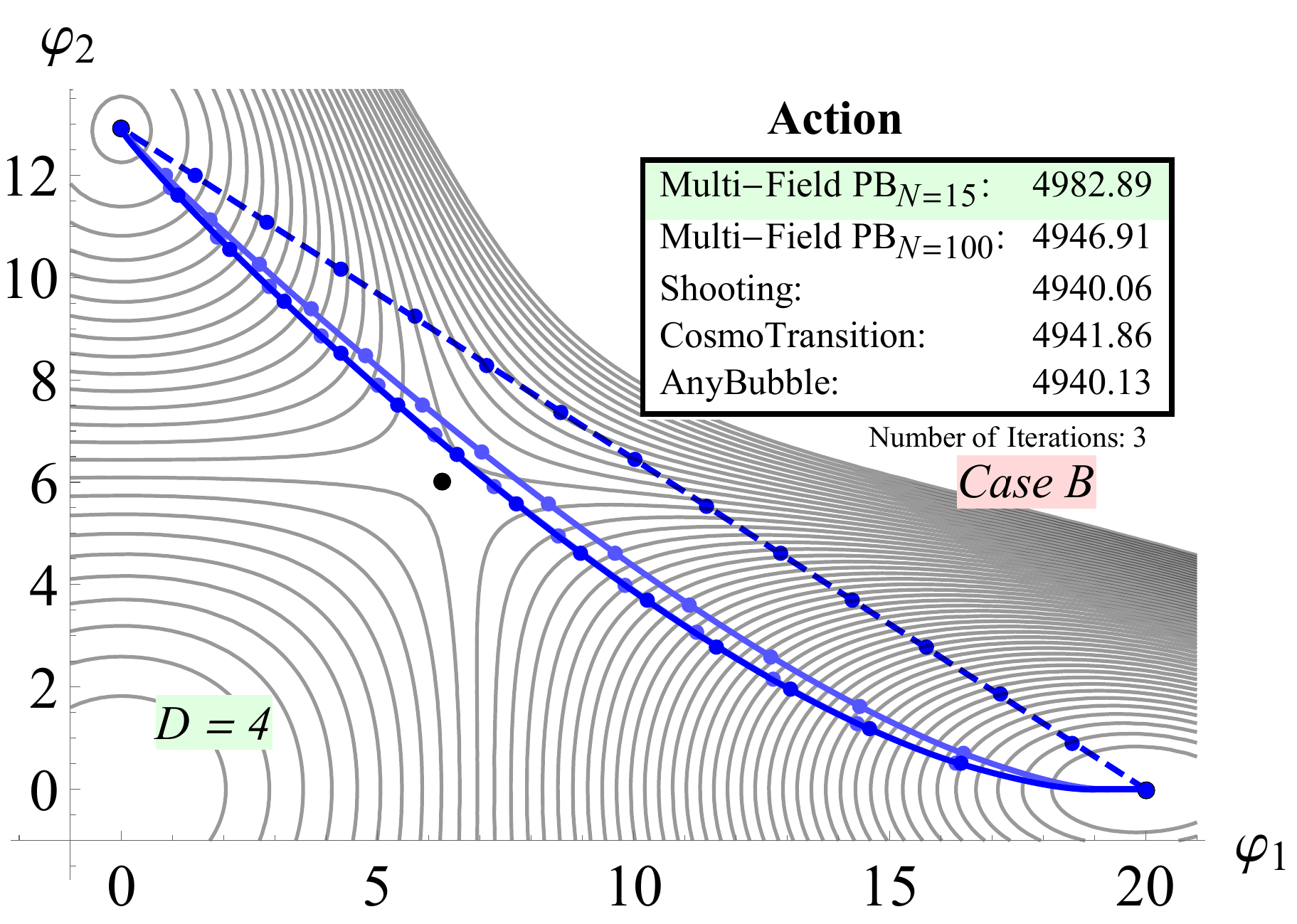} \hspace{1cm}
  \caption{Multi-field polygonal solution in $D=4$ with $N=15$ segmentation points. The starting ansatz is the straight
  dashed line connecting the two minima, shown as black dots, together with the saddle point. The solid
  lines are subsequent iterations that converge to the final path that solves the bounce equations. Insets show the
  action compared to other approaches.
  Left: The case a) set-up with the initial endpoint, which is free to move.
  Right: The case b) potential of the thin wall type with fixed endpoints in the minima.}
  \label{fig:mf_field_iterations}
\end{figure*}

The PB solution overcomes a number of these shortcomings and provides a framework with the following
features.
\begin{enumerate}
  \item The multifield PB field solution remains as simple as in the single field case in~\eqref{eqPBSol}.
  It is therefore fast to evaluate numerically and is retained upon iteration. The final result has
  a closed analytical form, which allows for further manipulation.

  \item The solution is built iteratively, where a single iteration takes into account the curvature
  in field space by explicitly solving the $\rho$ dependence and simultaneously deforms the path. This
  eliminates the oscillatory behavior and the solution converges quickly, within $\mathcal O(1)$ iterations,
  see FIG.~\ref{fig:mf_field_iterations}.

  \item The method works very well in the thin wall limit, which is usually problematic due to severe fine-tuning.
  This feature is directly inherited from the single field case and is due to the fact that we are solving for the
  Euclidean time $\rho$ variable and not in $\varphi$ space. Of course, the method works equally well (see again
  FIG.~\ref{fig:mf_field_iterations}) in the thick wall regime; moreover it is applicable to cuspy and unstable potentials,
  as well as paths with multiple minima.

  \item Finding the path in field space boils down to a coupled system of ordinary linear equations that
  scales linearly with the number of fields and number of segments. The procedure converges very close to the
  final path even with a few - $\mathcal O(1)$ segments. One can switch to more segments in the final step only
  to ensure sufficient precision in the longitudinal direction, depending on the desired precision of the action.

  \item It works for any space-time dimensions $D>2$ (with $D=2$ in the Appendix~\eqref{appD268}), in particular
  it is simple to consider $D = 3, 4$, which are most relevant for physical applications.
 \end{enumerate}

%
%
\subsection{Constructing multi-field polygonal bounces}

\begin{figure}[!h]
  \centering
  \includegraphics[width=.495\columnwidth]{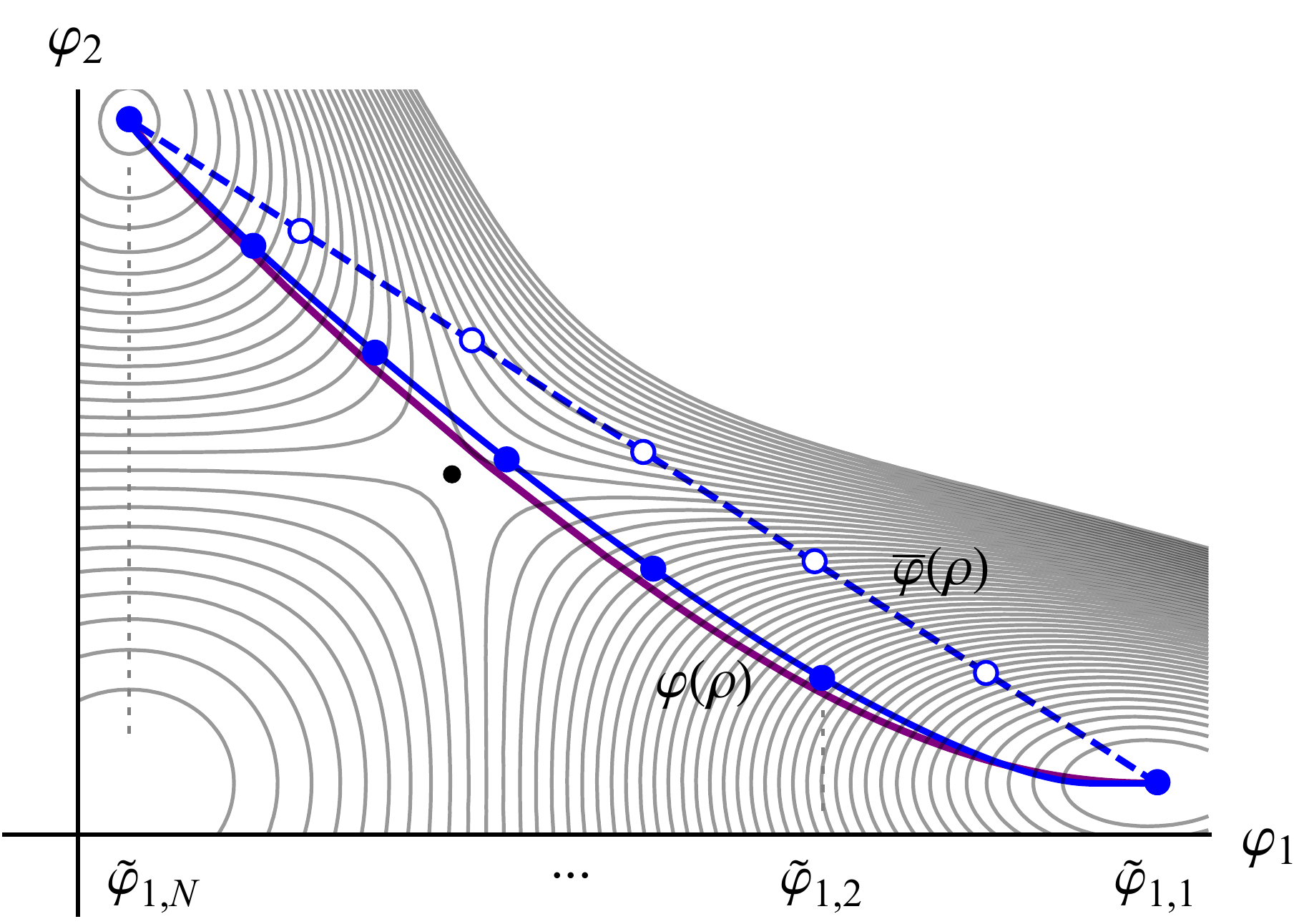}%
  \includegraphics[width=.495\columnwidth]{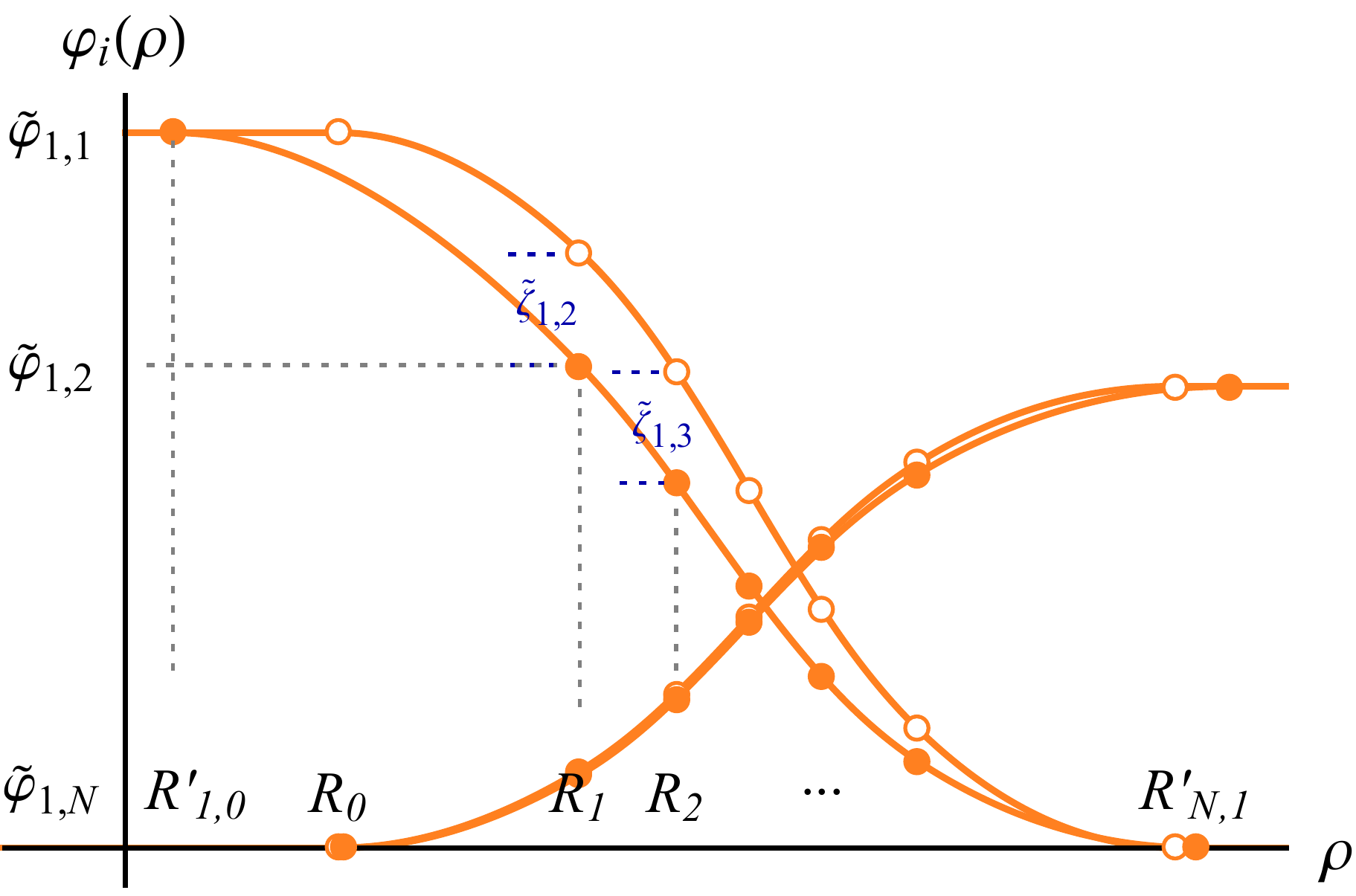}
  \caption{The PB solution for two fields in $D=4$ with $N=7$ segment points.
  Left: Path in field space with the initial straight line ansatz $\bar \varphi$ with empty
  circles and the first iteration of the PB solution in solid blue and full circles; the
  result from shooting is shown in purple.
  Right: Iterations of the evolution in Euclidean time for $\varphi_1(\rho)$.}
  \label{figMultiFieldBounce}
\end{figure}

Let us describe the generalization of the PB approach to an arbitrary number of scalar fields. The starting point is a single
field PB solution $\bar \varphi_{i s}$ where $i$ is the field index $i=1,\ldots,n_f$ and $s = 1,\ldots,N$
is the segment point. The ansatz is obtained from a selection of initial points in the multi-field space $\tilde \varphi_{i s}$,
for instance by segmenting a straight line connecting the two minima, as in the left panel of FIG.~\ref{figMultiFieldBounce},
and computing the corresponding longitudinal PB, seen on the right panel of FIG.~\ref{figMultiFieldBounce}.

We then consider an expansion around the initial estimate, such that $\varphi_{i s}(\rho) = \bar \varphi_{i s} +
\zeta_{i s}$. This produces a set of coupled bounce equations for each field direction
\begin{equation}
\begin{split} \label{eqColPhi}
  &\underbrace{\ddot {\bar \varphi}_{i s} + \frac{D-1}{\rho} \dot{\bar \varphi}_{i s}}_{8 \bar a_{i s}} +
  \underbrace{\ddot \zeta_{i s} + \frac{D-1}{\rho} \dot \zeta_{i s}}_{8 a_{i s}} = \frac{d V}{d \varphi_i} \left(\bar \varphi + \zeta \right).
\end{split}
\end{equation}
The idea here is to look for a solution of the field expansion $\zeta$ which is of the polygonal type
\begin{align} \label{eqPBZeta}
   \zeta_{i s} &= v_{i s} + \frac{2}{D-2} \frac{b_{i s}}{\rho^{D-2}} + \frac{4}{D} a_{i s} \rho^2,
\end{align}
where $a_{i s}$ corresponds to the leading constant expansion of the gradient of the potential
around some deformed path, defined by $\tilde \varphi_{i s} + \tilde \zeta_{i s}$.
This is the main difference in contrast to the single field case: the position in field space is not
fixed a priori and one has to allow for the segmentation to move in field space.

The gradient parameters $a_{i s}$ can be linearized in terms of the displacement $\tilde \zeta_{j s}$
with a symmetric average
\begin{align} \label{eqExpDer}
  8 a_{i s} &\simeq \frac{d V}{d \varphi_i}\left( \tilde \varphi_{i s} + \tilde \zeta_{i s} \right) - 8 \bar a_{i s},
  \\ \label{eqExpDer2}
  \frac{d V}{d \varphi_i} &\simeq \frac{d_i \tilde V_s + d_i \tilde V_{s+1} +
  d^2_{ij}\tilde V_s \tilde \zeta_{j s} + d^2_{ij} \tilde V_{s+1} \tilde \zeta_{j s+1}}{2}.
\end{align}
It is crucial that the gradient in~\eqref{eqExpDer2} is expanded beyond the constant leading order
up to $\mathcal O(\tilde \zeta)$ that includes the second derivative of the potential.
This is needed to properly describe curved paths in field space.

{\em Matching.} To fix the remaining parameters of the $\zeta$ solution in~\eqref{eqPBZeta}, the field
has to match onto the deformed path. We choose to match to $\tilde \zeta$ at the fixed radii $R_s$, computed
from the initial longitudinal polygonal ansatz. This can be done for all the $R_s$, except for the initial
$R_{i0}$ and final ones $R_{i N-1}$, which are free parameters for each field direction $i$.

The field values of the ansatz $\bar \varphi_{i s}$ are continuous from one section to another, while
the derivatives may not be. The matching of derivatives at $R_s$ then gives the recursion relation
for $b_{is}$
\begin{align} \label{eqRecRelBeta}
\begin{split}
  b_{i s} = b_{i 1} & + \sum_{\sigma = 1}^{s-1} \frac{4}{D} \left(a_{i \sigma+1} - a_{i \sigma} \right) R_\sigma^D
  \\
  & + \frac{1}{2} \left( \dot{\bar \varphi}_{i \sigma+1} - \dot{\bar \varphi}_{i \sigma} \right) R_\sigma^{D-1},
\end{split}
\end{align}
and field continuity, together with~\eqref{eqRecRelBeta} provides the recursion relation for $v_{is}$
\begin{align} \label{eqRecRelNu}
\begin{split}
  v_{i s} = v_{i 1} &- \sum_{\sigma = 1}^{s-1}  \frac{4}{D-2} \left(a_{i \sigma+1} - a_{i \sigma} \right) R_\sigma^2
  \\
  &- \frac{1}{D - 2} \left( \dot {\bar \varphi}_{i \sigma+1} - \dot {\bar \varphi}_{i \sigma} \right) R_\sigma.
\end{split}
\end{align}

{\em Initital/final conditions.}
In case a) the initial endpoint is free to move, however the solution
starts at $\rho = R_{i0} = 0$ with a vanishing derivative, therefore
\begin{align}
  v_{i1} &= \tilde \zeta_{i1}, & b_{i1} &= 0.
\end{align}
In case b) the initial endpoint does not move and we have
$\varphi_{i1} \left( R_{i0} \right) \simeq \bar \varphi_{i1} + {\dot {\bar \varphi}}_{i1} R_0 r_{i0} + \zeta_{i1} = \tilde \varphi_{i1}$
that implies $\zeta_{i1} (R_0) = \tilde \zeta_{i 1} = 0$ because ${\dot {\bar \varphi}}_{i1}(R_0) = 0$.
Here we expanded the initial and final radii $R_{i0} = R_0 \left(1+ r_{i0} \right)$ and $
R_{i N-1} = R_{N-1} \left(1+ r_{i N-1} \right)$ to leading order in $r_{i0,N-1}$, in order to maintain a linear system.
As for the derivatives,
\begin{align}
\begin{split}
  \dot \varphi_{i1} \left( R_{i0} \right) &\simeq {\dot {\bar \varphi}}_{i1} + {\ddot {\bar \varphi}}_{i1} R_0 r_{i0} + \dot \zeta_{i1}
  \\
  &= 8 \bar a_{i 1} R_0 r_{i0} + \dot \zeta_{i1} = 0,
\end{split}
\end{align}
where ${\dot {\bar \varphi}}_{i1} = 0$ and ${\ddot {\bar \varphi}}_{i1} = 8 \bar a_{i 1}$ follows from~\eqref{eqColPhi}.

In summary we have the following conditions
\begin{align} \label{eqZetaFinal}
  \zeta_{i1}(R_0) &= \zeta_{iN-1}(R_{N-1})  = 0,
  \\ \label{eqZetaDotInit}
  \dot \zeta_{i1}(R_0) &= - 8 \bar a_{i 1} R_0 r_{i0},
  \\ \label{eqZetaDotFinal}
  \dot \zeta_{iN-1}(R_{N-1}) &= - 8 \bar a_{i N-1} R_{N-1} r_{iN-1}.
\end{align}

The final task is to solve this linear system. The initial conditions are solved in terms of $v_{i1}$ and $b_{i1}$
\begin{align} \label{eqInitnu1beta1}
  v_{i1} &= -\frac{4}{D-2} \left(a_{i1} + 2 \bar a_{i1} r_{i0} \right) R_0^2,
  \\
  b_{i1} &= \frac{4}{D} \left( a_{i1} + D \bar a_{i1} r_{i0} \right) R_0^D,
\end{align}
which determines $\zeta_{i 1}$ that has to be fixed to $\tilde \zeta_{i 2}$ at $R_1$. The recursion relations~\eqref{eqRecRelBeta}
and~\eqref{eqRecRelNu} then provide the polygonal ansatz for $\zeta_{i s}$, to be fixed onto $\tilde \zeta_{i s+1}$
\begin{align} \label{eqFixZetaT}
  \zeta_{i s}(R_s) = \tilde \zeta_{i s+1}.
\end{align}
This continues until the final segment where the endpoint does not move anymore $\tilde \zeta_{i N} = 0$,
in agreement with~\eqref{eqZetaFinal}. The final equation to be solved is then the $\dot \zeta_{iN-1}$ condition in~\eqref{eqZetaDotFinal}.

By construction,~\eqref{eqPBZeta} keeps the same polygonal form in $\rho$,
therefore it is simple to iterate and converges once the path in field space does not change
anymore, i.e. $\tilde \zeta_{is} \simeq 0$.

%
%
\subsection{Examples and path convergence}
Let us consider a simple two field potential
\begin{align}
  V\left( \varphi_i \right) &= \sum_{i=1}^2 \left( -\mu_i^2 \varphi_i^2 + \lambda_i^2 \varphi_i^4 \right) +
  \lambda_{12} \varphi_1^2 \varphi_2^2 + \tilde \mu^3 \varphi_2,
\end{align}
that has multiple solutions for spontaneous symmetry breaking vevs $\langle \varphi_i \rangle = v_i$.
The metastable minima are in general of different depths with $V(v_1) \neq V(v_2)$, which allows for
the local false vacuum to decay into the global minimum by traversing the field space along the bounce solution.

To illustrate the multi-field PB method, we choose two exemplary points in the parameter space to cover
both non-trivial cases: a) and b). Specifically, we take $\mu_1^2 = 80$, $\mu_2^2 = 100$,
$\lambda_1 = 0.1$, $\lambda_2 = 0.3$, $\lambda_{12} = 2$ and $\tilde \mu^3  =800$ for case a), while $\tilde \mu=0$ for case b).
The solution in field space is shown on FIG.~\ref{fig:mf_field_iterations}, with the initial ansatz taken to be a
straight line with $N=15$ that connects the two minima. Remarkably, the PB solution converges to the
correct value very quickly, with $\mathcal O(1)$ iterations, as seen from FIG.~\ref{fig:mf_field_iterations}.

It is clear from the insets of FIG.~\ref{fig:mf_field_iterations} that the PB action is quite precise even with $N=15$
and reaches roughly permille precision with $N=100$. The main requirement for improving the precision of the action
is to increase the number of segments to get an accurate description of the longitudinal $\rho$ dependence. The
shape of the path in field space is less important and does not change much when $N$ increases. All of the
results above are similar for $D=3$.

Again, the convergence of the action can be improved by taking into account also the $\rho$ dependence of the
PB ansatz, similar to the single field extension defined above. It is also possible to solve the multifield bounce equation
by solving for $\zeta$ dynamically and gluing the corresponding Bessel functions. This is a somewhat tedious task that
requires local field rotations and is beyond the scope of the current work, but a similar semi-numerical approach was
done in $D=3$ by~\cite{Akula:2016gpl}.

Finally, the path converges to the final one without oscillations, in contrast to~\cite{Wainwright:2011kj} where
the $\rho$ dependence of transverse field directions was dropped, effectively neglecting the kinetic term.
Since we use an explicit solution in~\eqref{eqPBZeta}, the dynamical term of the curved path is taken
into account. This happens also in~\cite{Akula:2016gpl}, where the field construction is slightly more involved,
requiring local rotations and evaluation of Bessel functions.

%
%
\section{Conclusions and outlook} \label{secConclude}

An efficient and fast approach for calculating the false vacuum tunneling rate is developed
for arbitrary potentials with any number of fields up to the desired precision.
The method is based on the simple, well-known exact solution~\cite{Duncan:1992ai} that is
extended to any number of segments, space-time dimensions and number of scalar fields.

Usually, the simple single field problem of finding the bounce is solved by shooting - numerically
integrating the bounce equation and looking for the correct initial condition. Here instead, the
differential equations are solved exactly and are glued into a single continuously differentiable field.
The boundary conditions can be solved exactly and the field solution is computed recursively.
The remaining initial/final conditions are highly non-linear but can be solved by iterative use of
Derrick's theorem or numerical root finding.

In contrast to numerical integration, the PB solution is given by segmented polynomials.
This allowed for simple analytical manipulation, such as including corrections of higher orders in the
potential expansion, quantum or thermal fluctuations, expanding to more fields and it ultimately
reduces the computational cost. Because the one field solution depends on a single dimensional
parameter, which is the initial radius defined on some initial segment, the fine-tuning of initial conditions
is avoided. This is advantageous especially in the thin wall regime, where the usual shooting procedure struggles.

The method was applied to a number of single field examples, from the simplest displaced quartic potential to
more involved cases, such as the bi-quartic potential. The resulting bounce action converges quickly
with \(N \gtrsim \mathcal O(10)\) and reaches a permille level precision as seen in FIG.~\ref{fig:Ss}, where
the comparison with existing tools is made. The semiclassical bounce solution was also employed
in the calculation of the one-loop quantum corrections, i.e. the prefactor of the decay rate.

The simplest polygonal potential can serve as an ansatz to be perturbatively deformed in order to describe
the remaining non-linearities. These are generically important close to the extrema and their inclusion
improves the convergence of the bounce action, as seen from FIG.~\ref{fig:SNLO}.

The ability of perturbative expansion allows for the generalization to the multi-field case. The main challenge
with respect to the single field case is finding the path in field space. The PB approach solves it by
starting from an initial polygonal ansatz that is iteratively deformed by solving the bounce equations
at the leading order. Path deformation is solved by a linear system and converges very quickly without
oscillations such that the action is recovered to arbitrary precision within a few iterations.

In summary, we find that the PB method is a robust, precise and reliable way of
computing the semi-classical tunneling rate for any given potential.
This approach describes the false vacuum decay in flat space time, however the solution can also be used in curved
space-time within a small gravitational field approximation~\cite{Coleman:1980aw, Isidori:2007vm}.
The PB solution and its extension can thus provide a tool with an analytical insight in characterizing stable vacua
of theories with multiples scalar fields~\cite{Claudson:1983et, Moreno:1998bq, Huber:2006wf, Johnson:2008kc,
Greene:2013ida, Dine:2015ioa}, describing bubble nucleation and the quality of potential first order
phase transitions as well as the related spectrum of gravitational waves.

%
%
\acknowledgments
We would like to thank Borut Bajc for valuable discussions and suggestions.
The work of VG was supported by the Slovenian Research Agency's young researcher program under
the grant No. PR-07582. AM was partially supported by the H2020 CSA Twinning project
No. 692194, ''RBI-T-WINNING''. MN was supported by the Slovenian Research Agency under the
research core funding No. P1-0035 and in part by the research grant J1-8137.
MN acknowledges the support of the COST actions CA15108 - ``Connecting insights in fundamental physics''
and CA16201 - ``Unraveling new physics at the LHC through the precision frontier''.
%
%
\appendix

%
%
\section{On \(D=2, 6, 8 \) dimensions} \label{appD268}
Here we complete the treatment of the polygonal bounce construction for dimensions other
than \(D=3,4\), starting with the special instance of \(D=2\). The field solution is
\begin{equation}
  \quad \varphi_s(\rho) = v_s + 2 a_s \rho^2 - b_s \ln \rho^2.
\end{equation}
The \(b_1\) expression in~\eqref{eqInitCondDCaseB} remains the same, while \(v_1\) is obtained
from~\eqref{eqInitCondDCaseB} by replacing
\begin{equation} \label{eqRepD2}
    \frac{4}{D-2} R_s^2 \xrightarrow{D \to 2} 2 R_s^2 \left( 1 - \ln R_s^2 \right).
\end{equation}
Likewise the expression for the final condition of \(b_{N-1}\) in~\eqref{eqbvFinalDg2} remains the same,
and the same replacement of~\eqref{eqRepD2} should be used to obtain \(v_{N-1}\). The resulting action is
\begin{equation}
  \label{eqSEFinalD2}
  \begin{split}
  &S_2 = \pi R_0^2 \left( \tilde V_1 - \tilde V_N \right) + 2 \pi
  \sum_{s=1}^{N-1} \Biggl[ 6 a_s^2 \rho^4 + b_s^2 \ln(\rho^2) +
  \\
  &\frac{\rho^2}{2} \left(8 a_s(v_s - \tilde \varphi_s) + \tilde V_s - \tilde V_N - 8 a_s b_s \ln(\rho^2) \right) \Biggr]_{R_{s-1}}^{R_s},
\end{split}
\end{equation}
The matching conditions for \(D=2\) are slightly different
\begin{align}
  & v_s + 2 a_s R_s^2 - b_s \ln R_s^2 = \tilde \varphi_{s+1},\\
 & v_{s+1} + 2 a_{s+1} R_s^2 - b_{s+1} \ln R_s^2 = \tilde \varphi_{s+1},
  \\
 & 2 \left(a_{s+1} - a_s \right) R_s^2 + b_s - b_{s+1} = 0,
\end{align}
and the recursion relations in~\eqref{eqvbnDg2} are modified by applying the replacement of~\eqref{eqRepD2}
to \(v_s\). The radii in two dimensions are solved by
\begin{align} \label{eqRnD2}
  \quad &R_s^2 = -\frac{b_s}{2 a_s} W \left( - 2 \frac{a_s}{b_s} \exp \left(\frac{v_s - \tilde \varphi_{s+1}}{b_s} \right) \right),
\end{align}
where \(W(z)\) is the product log function that returns the solution of \(w\) to the equation \(z = w \, e^w\) for a
given \(z\).

The polygonal bounce setup for \(D=6,8\) closely follows the procedure outlined in~\ref{secConstruct} above,
apart from the solution of the radii fewnomial in~\eqref{eqRnDg2}. Indeed, the two closed form solutions
for \(D=6,8\) are
\begin{align}
  \begin{split} \label{eqRnD6}
  D = 6: \quad 2 R_s^2 &= \frac{\delta_s}{a_s} + \left(\frac{\delta_s}{a_s} \right)^2 \frac{1}{\zeta} + \zeta,
  \\
  \quad \zeta^3 &= \sqrt 3 \sqrt{3 \left(\frac{b_s}{a_s}\right)^2 - 2 \left(\frac{\delta_s}{a_s}\right)^3 \frac{b_s}{a_s}} + \\
  & \left(\frac{\delta_s}{a_s}\right)^3 - 3 \frac{b_s}{a_s},
  \end{split}
  \\
  \begin{split}\label{eqRnD8}
  D = 8: \quad 2 R_s^2 &= \frac{\delta_s}{a_s} - \chi_1 -\\
   &\sqrt{2 \left(\frac{\delta_s}{a_s}\right)^2 - \frac{2 \delta_s^3}{a_s^3 \chi_1} - \frac{\sqrt[3] 2}{3} \frac{4 b_s + \sqrt[3]2 a_s \chi_0^2}{a_s \chi_0}},
  \\
  \chi_1^2 &= \left(\frac{\delta_s}{a_s}\right)^2 + \frac{4 \sqrt[3]{2}}{3} \frac{b_s}{a_s \chi_0} + \frac{\sqrt[3]{4}}{3} \chi_0,
  \\
  \quad \chi_0^3 &= \sqrt{81 \left(\frac{\delta_s}{a_s}\right)^4 \left(\frac{b_s}{a_s}\right)^2 - 32 \left(\frac{b_s}{a_s}\right)^3} +\\
  & 9 \left(\frac{\delta_s}{a_s}\right)^2 \frac{b_s}{a_s}.
  \end{split}
\end{align}

%
%
\section{\(N = 3\) in \(D\) dimensions} \label{appN3D}
{\em Single-field.} The simplicity of having only three points allows for some further progress. In particular, the shooting in
\(\varphi_0\) for case a) can be carried out analytically for any \(D\).
%

{\em a)} Here, \(b_1 = 0\) and \(R_1\) is easy to solve from~\eqref{eqMatchField1Dg2}, while \(R_2\) follows
from~\eqref{eqMatchbN}
\begin{align}
  R_1^2 &=\frac{D}{4} \left(\frac{\tilde \varphi_2 - \varphi_0}{a_1}\right),
  &
  R_2^2 &= R_1^2 \left( \frac{a_2 - a_1}{a_2} \right)^{\frac{2}{D}}.
\end{align}
The recursion for \(v_s\) in~\eqref{eqvbnDg2} and the final condition for \(v_2\)~\eqref{eqbvFinalDg2} for \( N=3 \) give
\begin{align}
  \begin{split}
  v_2 &= \varphi_0 - \frac{4}{D-2} \left(a_2 - a_1 \right) R_1^2
  \\
  &= \tilde \varphi_3 - \frac{4}{D-2} a_2 R_2^2,
  \end{split}
  \\
  \varphi_0 &= \frac{\tilde \varphi_3 + c \, \tilde \varphi_2}{1 + c},
  \\
   & c = \frac{D}{D-2}\frac{a_2 - a_1}{a_1} \left(1 - \left(\frac{a_2}{a_2 - a_1} \right)^{\frac{D-2}{D}} \right).
\end{align}
For case a) to be consistent, the final solution should obey \(\varphi_0 > \tilde \varphi_1\).

%
%
{\em b)}
Plugging the initial/final conditions of~\eqref{eqInitCondDCaseB} and \eqref{eqbvFinalDg2} into~\eqref{eqMatchbN},
\eqref{eqMatchField1Dg2} and~\eqref{eqvbnDg2} gives
\begin{align}
  a_2 \left( R_2^D - R_1^D \right) + a_1 \left(R_1^D - R_0^D \right) =0,
  \\
  R_1^2 + \frac{2}{D - 2} R_1^{2 - D} R_0^D - \frac{D}{D - 2} R_0^2 = \frac{\tilde \varphi_2 - \tilde \varphi_1}{a_1},
  \\
  \frac{4}{D - 2} \left(a_2 \left( R_2^2 - R_1^2 \right) + a_1 \left( R_1^2 - R_0^2 \right) \right) =  \tilde \varphi_3 - \tilde \varphi_1.
\end{align}
This system can be reduced to a single non-linear equation that can be solved numerically for any \(D\).
However, \(D = 4\) is special, here a simple closed form solution can be obtained. The above equations can be rewritten as
\begin{align} \label{eqMatchCondb}
  \left( R_1^2 - R_0^2 \right)^2 &= R_1^2 \left( \frac{\tilde \varphi_2 - \tilde \varphi_1}{a_1} \right) = R_1^2 \Delta_2^2,
  \\
  \left( R_2^2 - R_1^2 \right)^2 &= R_1^2 \left(\frac{\tilde \varphi_3 - \tilde \varphi_2}{-a_2} \right) = R_1^2 \Delta_3^2,
  \\ \label{eqMathDrCondb}
  a_1 \left( R_0^4 - R_1^4 \right) &= a_2 \left( R_2^4 - R_1^4 \right).
\end{align}
Expressing \(R_0^2 = R_1 \left( R_1 - \Delta_2 \right)\) and \(R_2^2 = R_1 \left( R_1 + \Delta_3 \right)\) and
plugging \(R_{0,2}\) into~\eqref{eqMathDrCondb} gives
\begin{equation}
\begin{split}
  R_1 &= \frac{1}{2} \frac{a_1 \Delta_2^2 - a_2 \Delta_3^2}{a_1 \Delta_2 + a_2 \Delta_3} \\
  &=  \frac{1}{2} \frac{\tilde \varphi_3 - \tilde \varphi_1}{\sqrt{a_1 \left( \tilde \varphi_2 - \tilde \varphi_1 \right)} -
   \sqrt{-a_2 \left( \tilde \varphi_3 - \tilde \varphi_2 \right)}} > 0.
\end{split}
\end{equation}

\begin{figure}[!ht]
  \includegraphics[width=.49\columnwidth]{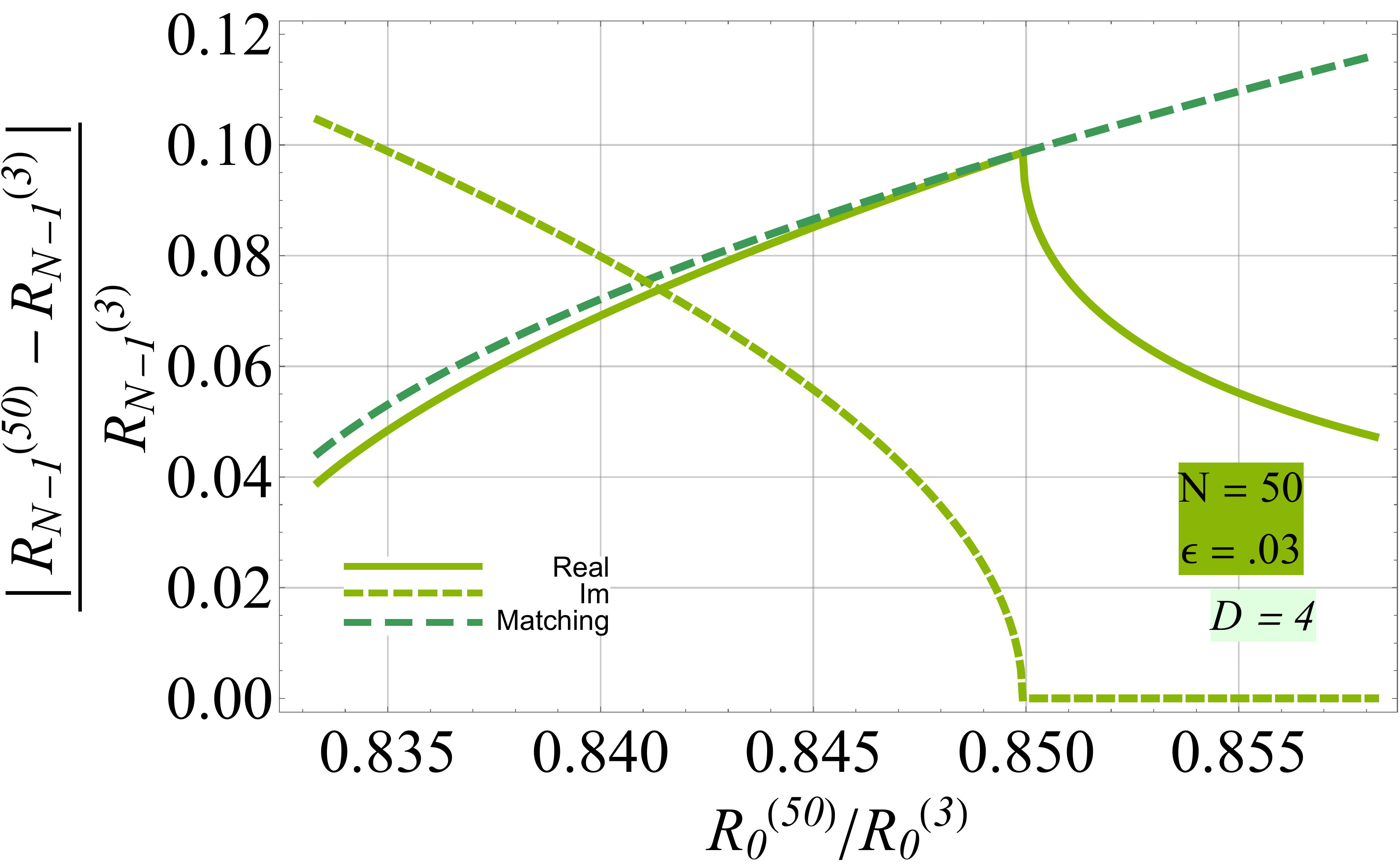}%
  \includegraphics[width=.49\columnwidth]{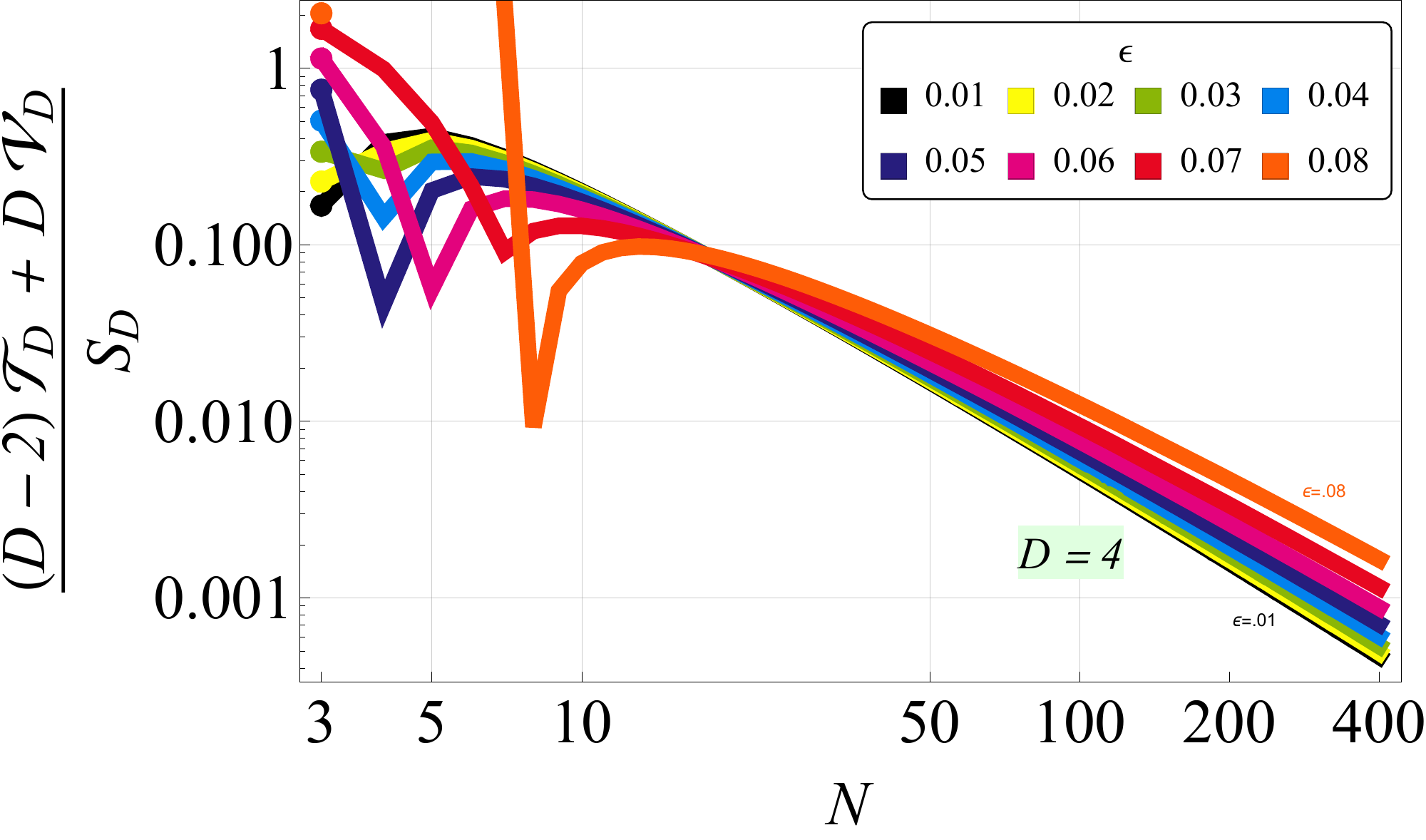}
  \caption{Left: The final radius dependence on \(R_0\) for \(N=50\) and \(\varepsilon = 0.03\), showing the
  real and imaginary part, as well as the corresponding value obtained from \(R_{N-1}\) in the matching
  condition in~\eqref{eqMatchbN}.
  Right: The continuous version of Derrick's theorem~\eqref{eqSDerrickCont} with \(\mathcal T\) computed with the
  PB and \(\mathcal V\) from the input potential in~\eqref{eqVC}. The normalized quantity acts
  as a test of convergence and goodness of approximation.}
  \label{fig:RfDk}
\end{figure}

%
%
{\em Multi-fields.} The minimal multi-field case with $N=3$ can be carried out analytically up to a single
$n_f^2$ linear system. The initial conditions in~\eqref{eqZetaFinal} and~\eqref{eqZetaDotInit}
with recursion relations~\eqref{eqRecRelBeta} and~\eqref{eqRecRelNu} give
\begin{align}
  \begin{split}
  v_{i2} &= -\frac{4}{D-2} \bigl(\left(a_{i1} + 2 \bar a_{i1} r_{i0} \right) R_0^2 +
  \\
  &\left(a_{i2} - a_{i1} \right) R_1^2 \bigr) - \frac{1}{D - 2} \left( \dot {\bar \varphi}_{i 2} - \dot {\bar \varphi}_{i 1} \right) R_1,
  \end{split}
  \\
  \begin{split}
  b_{i2} &= \frac{4}{D} \bigl( \left( a_{i1} + D \bar a_{i1} r_{i0} \right) R_0^D +
  \\
  &\left(a_{i2} - a_{i1} \right) R_1^D \bigr) +
  \frac{1}{2} \left( \dot {\bar \varphi}_{i 2} - \dot {\bar \varphi}_{i 1} \right) R_1^{D-1}.
  \end{split}
\end{align}
This leaves us with three equations for $r_{i 0}, r_{i 2}$ and $\tilde \zeta_{i 2}$
\begin{align} \label{eqri0N3}
  \begin{split}
  r_{i0} &= \biggl( \frac{D-2}{8} R_1^{D-2} \tilde \zeta_{i2} -
  a_{i 1} \bigl(\frac{D-2}{2 D} R_1^D
  \\
  & - \frac{R_0^2 R_1^{D-2}}{2} + D R_0^D \bigr) \biggr) /\bar a_{i1} \left(R_0^D - R_0^2 R_1^{D-2} \right) ,
  \end{split}
  \\ \label{eqzt2N3}
  v_{i 2} &+ \frac{2}{D-2} \frac{b_{i 2}}{R_2^{D-2}} + \frac{4}{D} a_{i 2} R_2^2 = 0,
  \\ \label{eqri2N3}
  r_{i2} &= \frac{1}{\bar a_{i 2}} \left(\frac{b_{i 2}}{4 R_2^{D}} - \frac{a_{i 2}}{D} \right),
\end{align}
Inserting $r_{i0}$ from~\eqref{eqri0N3} into~\eqref{eqzt2N3} gives a linear system for $\tilde \zeta_{i2}$
that can be solved using the explicit form of $a_{i 1,2}(\tilde \zeta_{i 2})$ given in~\eqref{eqExpDer}.
Once $\tilde \zeta_{i2}$ is given, $r_{i2}$ follows from~\eqref{eqri2N3}, which concludes the
calculation of $\zeta$.

Remarkably, this simple estimate already gives a rather good approximation for the path in field space,
the main inaccuracy in the bounce action is due to the poor estimate of the $\rho$ dependence.

%
%
\section{Real radii and root finding} \label{appRootFReR}

\paragraph{Real radii.}

The radii solutions in Eqs.~\eqref{eqRnD2}-\eqref{eqRnD8}, as well as those in~\eqref{eqRnD3},\eqref{eqRnD4}
above, allow for a number of branches. The ones chosen above are such that the resulting \(R_s\) are real
and positive. Moreover, the slope of the potential \(a_s\) has to be appropriately factorized in the expressions
above in order to maintain the reality of \(R_s\) during the transition through the maximum of \(V\) when
\(a_s\) flips the sign. This choice of signs also ensures that the radii of segments below the initial \(\varphi_0\)
automatically remain 0, i.e. \(R_s = 0\) for \(\tilde \varphi_s < \varphi_0\).

\paragraph{Root finding.}
The starting point for root finding is to determine the real domains of the initial parameters \(\varphi_0\)
and \(R_0\) for a) and b) cases, respectively. This defines the region of parameter space where a consistent
solution can be searched for. To illustrate this point, we show the behavior of the final radius with
respect to \(R_0\) and \(\varphi_0\) in FIG.~\ref{fig:RfDk}.
It is curious that the solution to the matching equation in~\eqref{eqMatchbN} lies precisely on the
edge of the real domain.

In order to implement the root searching numerically, one has to define a starting estimate for \(R_0\)
or \(\varphi_0\). It turns out that for case a) the more stable option is to choose the initial estimate
for \(\varphi_0\) close to the false vacuum \(\varphi_0 \simeq \tilde \varphi_1\), while in the case b)
the \(N=3\) result gives a fairly reliable starting point. Moreover, the behavior of case b) root finding
convergence is in general more stable with respect to case a).

\begin{figure}[t]
  \includegraphics[width=.49\columnwidth]{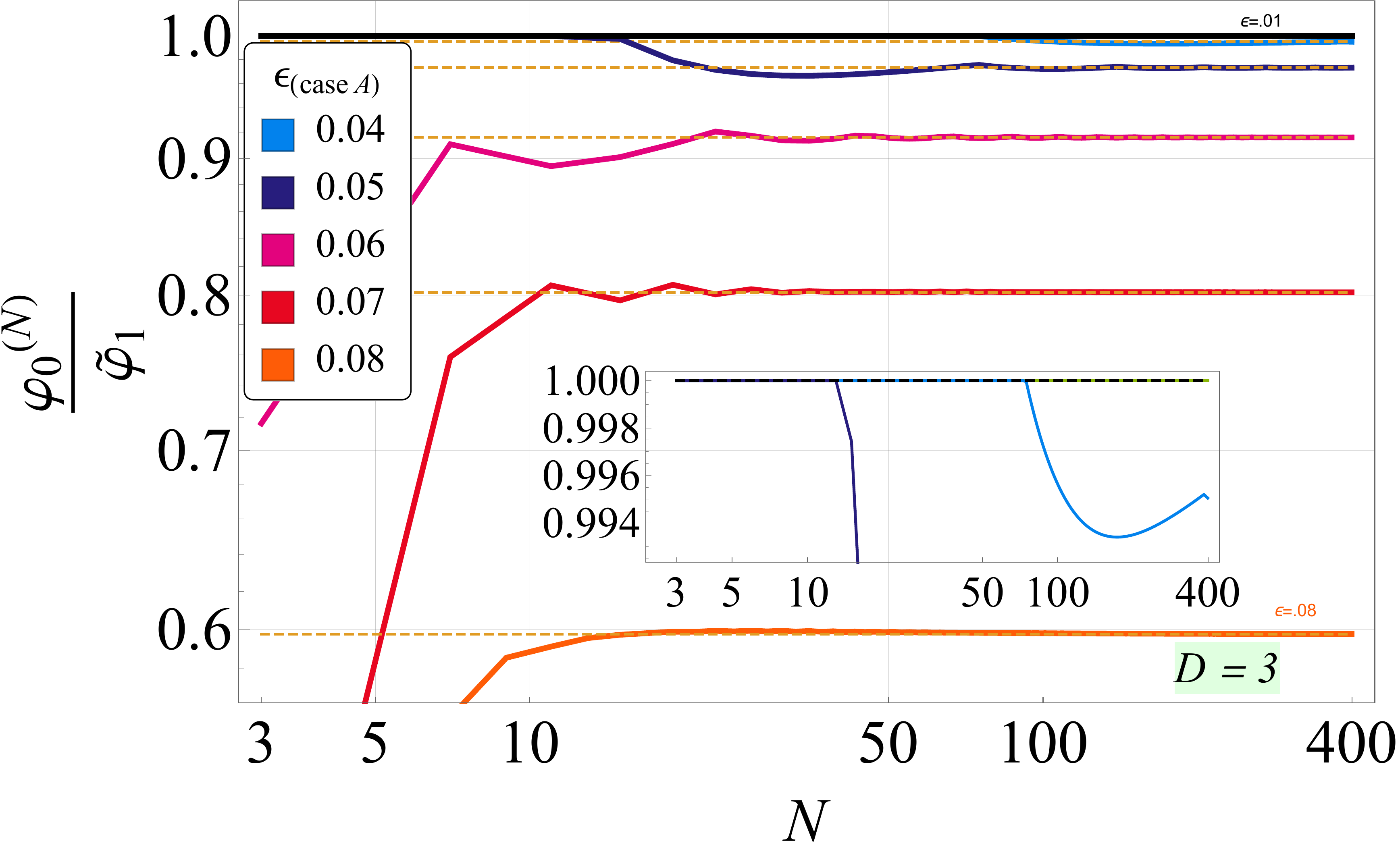} %
  \includegraphics[width=.49\columnwidth]{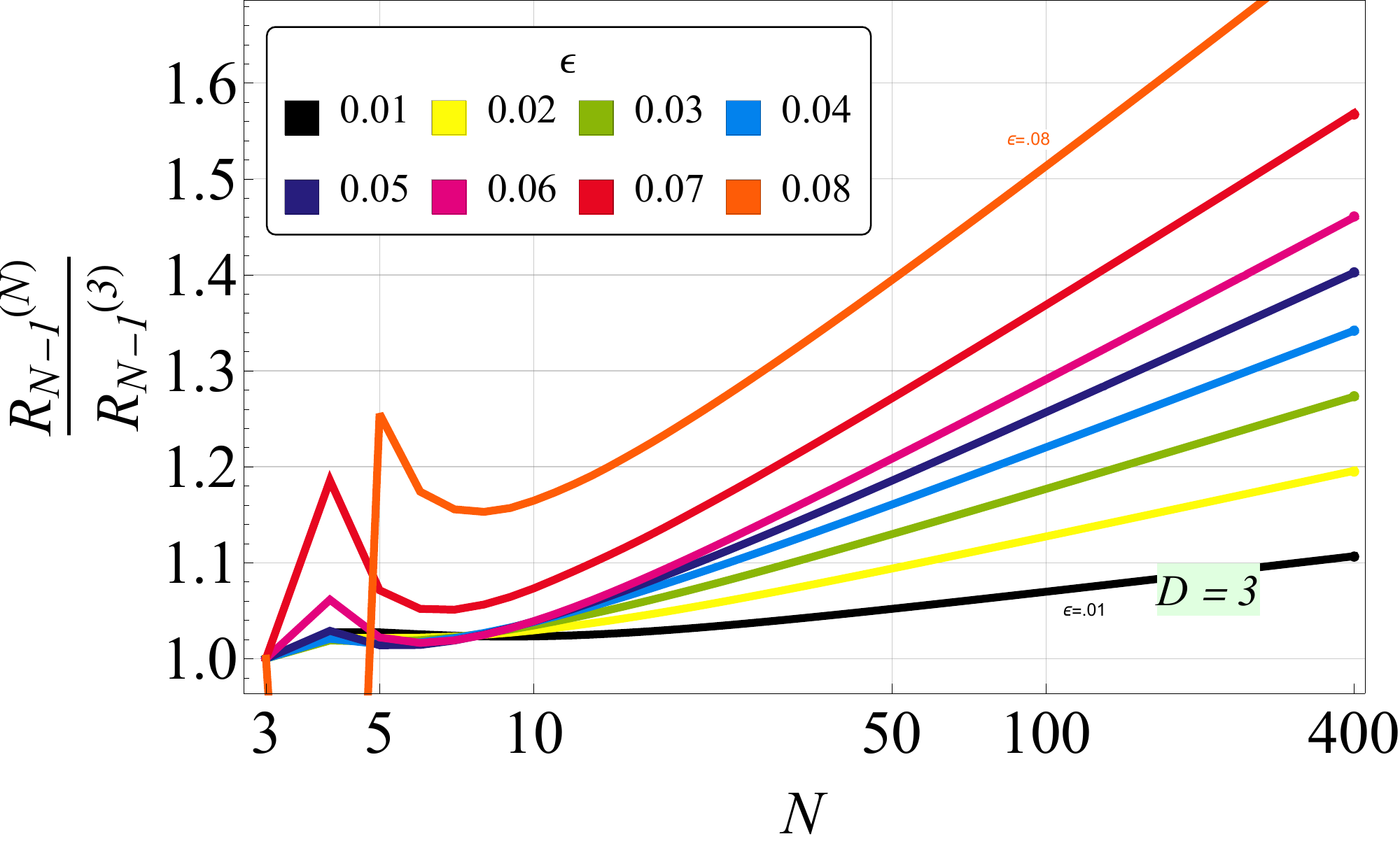}
  \caption{Left: The initial field value \(\varphi_0\) normalized to the position of the false minimum
  in \(\tilde \varphi_1\). Right: The final radius \(R_{N-1}\), normalized to the \(N=3\) approximation.}
  \label{fig:Phi0Rfs}
\end{figure}

The behavior of \(\varphi_0\) that solves the polygonal bounce in case a), is shown on the left of
FIG.~\ref{fig:Phi0Rfs}, where the field is normalized to the position of the false minimum
in \(\tilde \varphi_1\). Notice that as \(\varepsilon\) decreases, the solution gets closer
to \(\tilde \varphi_1\) and eventually crosses over to case b).
The smaller \(N\) approximation typically underestimates the final value and oscillates
towards the limiting value, which is an artefact of the segmentation.

Note also that for \(\varepsilon = 0.05(0.04)\), the solution for case a) does not exist
until \(N \gtrsim 10(70)\) when the segmentation becomes refined enough for the method
to work and which is precisely when \(R_0\) becomes non-zero in FIG.~\ref{fig:Rs}.
Another particularity related to the segmentation happens with \(\varepsilon = 0.07\) in
\(D=4\) where we start in case a) for \(N=3,4\), switch to case b) and return back to a)
at \(N=8\).

The right panel of FIG.~\ref{fig:Phi0Rfs} shows the extent of the non-trivial part of the
bounce field solution in the \(\rho\) dimension, i.e. the final radius \(R_{N-1}\), normalized to
the \(N=3\) approximation. Above this radius, the bounce solution remains constant as in
FIG.~\ref{fig:VPhi}. As we expect to get back to~\eqref{eqBounceEqD} in the continuous
limit, the \(R_{N-1}\) should go to infinity when \(N\) increases, which is evident from
the right panel of FIG.~\ref{fig:Phi0Rfs}.

As discussed above, the \(R_{N-1}\) is a finite and numerically well defined quantity that
regulates the infinity of \(\rho\). In particular, the extent to which the final radius grows is
surprisingly small. Even for a large number of
points \(N \sim 400\) where the bounce action is already quite precise, the final radius is
merely about 50\% larger than the initial estimate from \(N=3\).

\def\arxiv#1[#2]{\href{http://arxiv.org/abs/#1}{[#2]}}
\def\Arxiv#1[#2]{\href{http://arxiv.org/abs/#1}{#2}}

\end{document}